\documentclass[a4paper,twocolumn,11pt,unpublished]{quantumarticle}
\pdfoutput=1
\usepackage[utf8]{inputenc}
\usepackage[english]{babel}
\usepackage[T1]{fontenc}
\usepackage{amsmath}
\usepackage{hyperref}

\usepackage{amsmath}
\usepackage{amsfonts}
\usepackage{amssymb}
\usepackage{braket}
\usepackage[nocompress]{cite}
\usepackage{multirow}
\usepackage{subcaption}
\usepackage[nolist,nohyperlinks]{acronym}

\long\def\comment#1{}

\def\parah#1{\vspace*{0.0in} \noindent{\bf #1:}}

\newcommand{\re}{\mathrm{Re}}

\newcommand{\tr}{\mathrm{Tr}}
\newcommand{\cE}{\mathcal{E}}

\newcommand{\qfacsp}{$QFactor$ }
\newcommand{\qfac}{$QFactor$}
\newcommand{\qfacsa}{$QFactor$-$Sample$}
\newcommand{\qfacsasp}{$QFactor$-$Sample$ }
\newcommand{\balpha}{\boldsymbol{\alpha}}

\title{Leveraging Quantum Machine Learning Generalization to Significantly Speed-up Quantum Compilation}

\author{Alon Kukliansky}
\affiliation{Naval Postgraduate School, 1 University Circle, Monterey, California 93943, USA}
\email{alon.kukliansky.is@nps.edu}
\orcid{0009-0003-6743-9018}

\author{Lukasz Cincio}
\affiliation{Theoretical Division, Los Alamos National Laboratory, Los Alamos, NM 87545, USA}
\email{lcincio@lanl.gov}
\orcid{0000-0002-6758-4376}

\author{Ed Younis}
\affiliation{Applied Mathematics and Computational Research Division,
Lawrence Berkeley National Laboratory, Berkeley, California 94720, USA}
\orcid{0000-0002-1306-1860}
\email{edyounis@lbl.gov}

\author{Costin Iancu}
\affiliation{Applied Mathematics and Computational Research Division,
Lawrence Berkeley National Laboratory, Berkeley, California 94720, USA}
\orcid{0000-0001-7845-2427}
\email{cciancu@lbl.gov}

\begin{document}


\begin{acronym}
\acro{CPU}{central processing unit}
\acro{GPU}{graphical processing unit}
\acro{NPS}{Naval Postgraduate School}
\acro{OOM}{out of memory}
\acro{GPO}{general-purpose optimizer}
\acro{SVD}{singular value decomposition}
\acro{QoR}{quality of result}
\acro{QML}{quantum machine learning}
\acro{ML}{machine learning}

\maketitle

\begin{abstract}

Existing numerical optimizers deployed in quantum compilers use expensive $\mathcal{O}(4^n)$ matrix-matrix operations. Inspired by recent advances in quantum machine learning (QML), \qfacsasp replaces matrix-matrix operations with simpler $\mathcal{O}(2^n)$ circuit simulations on a set of sample inputs. The simpler the circuit, the lower the number of required input samples. We validate \qfacsasp on a large set of circuits and discuss its hyperparameter tuning.  When incorporated in the BQSKit quantum compiler and compared against a state-of-the-art domain-specific optimizer, we demonstrate improved scalability and a reduction in compile time, achieving an average speedup factor of {\bf 69} for circuits with more than 8 qubits. We also discuss how improved numerical optimization affects the dynamics of partitioning-based compilation schemes, which allow a trade-off between compilation speed and solution
quality.

\end{abstract}

\section{Introduction}

Given a parameterized quantum circuit and a target unitary, a common operation in quantum program development is to solve an optimization problem to determine the parameters that implement the target unitary. Solving for parameters is commonly referred to as \emph{instantiation}, and it is an operation that appears in hybrid algorithms~\cite{vqe, qaoa}, circuit synthesis~\cite{cpflow, squander1, nacl, bqskit, qfast,qsearch,qce22_ed} or within \ac{QML}~\cite{cerezoVariationalQuantumAlgorithms2021,benedettiParameterizedQuantumCircuits2019,biamonteQuantumMachineLearning2017,beerTrainingDeepQuantum2020} algorithms.

In all existing approaches, the objective function in instantiation requires computing process distances~\cite{qce22_ed} between two unitaries, an operation with $\mathcal{O}(4^n)$ complexity. As far as we know, the state-of-the-art is illustrated by the \qfac~\cite{kuklianskyQFactorDomainSpecificOptimizer2023a} domain-specific optimizer, which uses a tensor network formulation together with analytic methods and an iterative local optimization algorithm to reduce the \emph{effective} number of problem parameters. The improvements over \acp{GPO} come, among other features, from working at the unitary rather than the parameter level. A given gate may have a very complicated representation in terms of parameters that need to be resolved by \acp{GPO}. In contrast, \qfacsp optimizes that gate on each update.

\qfacsp improves performance and scalability by reducing the number of parameters by a (large) constant. In this paper, we show how to further gain $\mathcal{O}(2^n)$ speedup in computational complexity for the algorithm's inner loop, while maintaining the same \acp{QoR}. Our benchmarks show an average reduction of 17X in runtime and an impressive 69X average reduction for circuits with 9-12 qubits. We have seen a runtime reduction of up to 830X for individual instantiation runs.

The basic idea is taken from recent advances~\cite{abbasPowerQuantumNeural2021, caroGeneralizationQuantumMachine2022, banchiGeneralizationQuantumMachine2021, duEfficientMeasureExpressivity2022} in \ac{QML} theory. Modern \ac{QML} methods involve variationally optimizing a parameterized quantum circuit on a training dataset and subsequently making predictions on unseen data (i.e., generalizing). It has been shown that for a quantum circuit with T parametrized gates, that has been trained on M samples the generalization error is bounded by $\mathcal{O}\left(\sqrt{\frac{T\log T}{M}}\right)$. We use a reduction from the instantiation problem to a traditional \ac{QML} flow, linking the generalization error to the instantiation error and taking advantage of this bound to limit the size of the training set, reducing the overall complexity.

\qfacsasp is an instantiation algorithm using a \ac{QML} approach. Given a target unitary, it randomly draws $M$ orthogonal states and applies the unitary on them to create the training set. It then performs optimization based only on that set. It has an \textit{on-the-fly} mechanism that can increase the size of the training set if it detects that the generalization error is too big. The simpler the unitary, the fewer the training examples required. In contrast, other optimizers use the full target unitary, with $2^n$ states to perform the optimization.

We integrated \qfacsasp into the BQSKit~\cite{bqskit} synthesis infrastructure and evaluated its performance in optimizing large quantum circuits. Compared to the next best optimizer, our results show an average 4-9X reduction in runtime. Additionally, by increasing the partition size beyond the capability of other optimizers, we observe an improvement in \acp{QoR}, although this comes with the trade-off of increased runtime.

The rest of the paper is organized as follows: Section~\ref{sec:background} provides background on \qfacsp instantiation algorithm and how to utilize \ac{QML} generalization error bound to improve instantiation. \qfacsasp algorithm and its implementation are described in Section~\ref{sec:qfs_algorithm}. Our evaluation procedures and results are presented in Section~\ref{sec:qfs_evaluation}. Finally, in Section~\ref{sec:discussion} we discuss our results.

\section{Background}\label{sec:background}

\subsection{Numerical Optimization and Instantiation}

Given a $2^n \times 2^n$ unitary $U$ and a parametrized circuit $ C: \mathbb{R}^k \mapsto \mathcal{U}(2^n) $, instantiation finds parameters $\balpha$ such that some (e.g. Froebenius) norm between $U$ and $C(\balpha)$ is minimized. 
\begin{equation} \label{eq:qfactor_cost_func}
    || U - C(\balpha) ||^2 = 2^{n+1} \left( 1 - \tfrac{\re \tr \left[U^\dagger C(\balpha)\right]}{2^n} \right) .
\end{equation}

The heaviest part of any existing instantiation algorithm is determined by the $\mathcal{O}(4^n)$ computational complexity of the norm in \eqref{eq:qfactor_cost_func}  (or part of it ) calculation, where $n$ is the number of qubits in the circuit.

As far as we know, \qfac~\cite{kuklianskyQFactorDomainSpecificOptimizer2023a} is the state-of-the-art in numerical optimization for instantiation. It is a domain-specific optimizer that reduces circuit parameter complexity by directly updating (possibly multi-qubit) unitary gates without parameterizing them internally. By treating each gate as a unitary without parameterization during optimization, it effectively reduces the parameter space, distinguishing it from conventional \acp{GPO}.

\qfacsp utilizes tensor network contractions for efficient circuit manipulations. The algorithm sweeps through a circuit, traversing gate (tensor) by gate (tensor), and at each gate, it executes a local optimization process to update the gate's unitary. The basic step in the sweep is the calculation of $\cE$, the gate's environment matrix~\cite{orus2014practical,kuklianskyQFactorDomainSpecificOptimizer2023a}. The algorithm then performs a \ac{SVD} on $\cE=XDY^\dagger$ resulting in an optimized gate.
\begin{equation}\label{eq:opt_u}
u_{k_{new}}=YX^\dagger.
\end{equation}

When compared to \acp{GPO}~\cite{nocedal1980updating,liu1989limited,ranganathan2004levenberg}, 
\qfacsp scales better with circuit depth and qubit count, achieving faster runtimes and better success rates.

Numerical optimization can be scaled in two ways: by reducing the number of parameters needed to describe a problem, or by decreasing the computational effort required per parameter instantiation. \qfacsp enhances scalability through the former approach. A key question now is whether we can develop a more efficient optimization procedure, specifically by lowering the cost of the objective function. 

\subsection{Quantum Generalization Error Bound}

A variational quantum circuit $C(\balpha)$ is a circuit with parameters that are adjusted iteratively to optimize a specific objective function. If its parameters $\balpha$ are trained with respect to some loss function $l$ over a training set $S=\{\left(\ket{\psi_i}, \ket{\phi_i}\right)\}_{i=1}^M$ then the training loss or training error is usually defined as
\begin{equation}\label{eq:tr_err}
    l_{\text{train}}=\frac{1}{M} \sum_{i=1}^M l(C(\balpha)\ket{\psi_i}, \ket{\phi_i}).
\end{equation}
For a given set of parameters $\balpha$, the expected prediction error over some distribution of states $\xi$ is
\begin{equation} \label{eq:pred_err}
    l_{\text{pred}} = \underset{\ket{\psi}, \ket{\phi} \sim \xi}{\mathbb{E}}[l\left(C(\balpha)\ket{\psi}, \ket{\phi}\right)].
\end{equation}

The \emph{generalization error} is defined as the difference between the expected prediction error~\eqref{eq:pred_err} and the training error~\eqref{eq:tr_err}, when $\xi$ is the distribution the training set was drawn from:
\begin{equation}\label{eq:gen_err}
    \text{gen}(C(\balpha))=l_{\text{pred}} - l_{\text{train}}.
\end{equation}

There are many \ac{ML} algorithms that, in addition to the training set, utilize a validation set for various purposes such as early stopping of the training process when overfitting is detected, finding optimal hyperparameters of the underlying training algorithm, or estimating the generalization error of the model~\cite{aurelien2019hands}.

It has been proven in~\cite{caroGeneralizationQuantumMachine2022} that for a quantum model with $T$ parametrized gates that has been trained on $M$ samples, the generalization error is bounded by:
\begin{equation}\label{eq:gen_bound}
    \text{gen}(C(\balpha))\in \mathcal{O}\left(\sqrt{\tfrac{T \log{T}}{M}}\right).
\end{equation}

This means that for circuits with $n$ qubits, if their number of gates is polynomial with respect to $n$ (poly-n), one does not need an exponential number of training states to achieve low generalization error, but rather only poly-n training states will suffice. The class of circuits that can be efficiently implemented on a quantum computer also has poly-n gates, therefore we can effectively use the generalization bound to limit the number of states used during training, thus improving the training runtime performance. This reduction in runtime is expected to improve as the circuit has more qubits in the same way $\tfrac{\text{poly-$n$}}{\text{exp-$n$}}$ shrinks as $n$ increases.

\subsection{QML and Instantiation}

Quantum circuit instantiation can be reduced to a conventional \ac{QML} flow. For a given unitary $U$, we generate a training set $\{\left(\ket{\psi_i}, U\ket{\psi_i}\right)\}_{i=1}^M$ by first randomly selecting $M$ mutually orthogonal states $\{\ket{\psi_i}\}_{i=1}^M$ as input training states. Then, we apply the unitary $U$ to these states to obtain the output training states. Following, we define the loss function to be the distance square between the two states
\begin{equation}
    l(\ket{\psi}, \ket{\phi}) = ||\ket{\psi} - \ket{\phi} ||^2,
\end{equation}
and the training loss to be the average loss over the training set
\begin{equation}
    l_\text{train}(C,\balpha)=\frac{1}{M} \sum_{i=1}^M l(C(\balpha)\ket{\psi_i}, U\ket{\psi_i}).
\end{equation}

\section{Algorithm} \label{sec:qfs_algorithm}

\qfacsasp reduces \qfac's computational complexity from $O(4^n)$ to $O(M2^n)$, where $n$ is the number of qubits and $M$ is the number of states required for training, by making use of known bounds~\cite{caroGeneralizationQuantumMachine2022} on the generalization error. The algorithm finds the optimal circuit parameters for only a small set of training states, and the expected error on all the other possible input states is bounded by~\eqref{eq:gen_bound}. Although \qfacsasp and \qfacsp have different cost functions, their basic optimization step is the same, where they locally optimize each gate at a time, performing an \ac{SVD} on the gate environment matrix.

The algorithm begins by generating a random orthogonal set of training states, $\{\ket{\psi_j}\}_{j=1}^M$, sampled from the Haar random distribution. Then it sweeps the circuit from right to left. For each gate, it calculates the environment matrix $\cE$, performs an \ac{SVD} on $\cE$, and updates the gate using~\eqref{eq:opt_u}. The sweep is repeated several times until a stopping condition has been reached. Fig.~\ref{fig:env_calc_tensor} presents the $\cE$ calculation and the algorithm state during a sweep. A short discussion about the importance of the training states distribution can be seen in Appendix~\ref{app:tsd}.

Since we do not know upfront the number of training states required, the algorithm incorporates an 'on-the-fly' approach for estimating the generalization error. This is achieved by comparing the cost values for the training states with those of a randomly selected set of validation states. If the observed error surpasses a predefined threshold, the algorithm halts and restarts with double the number of training states. This iterative process terminates upon reaching the desired convergence threshold, detecting a plateau, or reaching the maximum limit of training states. We would like to point out that the \emph{double-and-restart} process doesn't change the asymptotic complexity, as $2+4+8+16+\cdots +m < 2m$.

In \qfac, the cost function~\eqref{eq:qfactor_cost_func} is the Froebenius norm between the target unitary and the instantiated circuit whereas in \qfacsa, the cost function we use is the average distance between the states generated by applying the target unitary on the training states and the states generated by applying the instantiated circuit on the same training states:
\begin{equation} \begin{split} \label{eq:qfactor_sample_cost_function}
    \frac{1}{M} \sum_{j=1}^M &||U\ket{\psi_j}-C(\balpha)\ket{\psi_j}||^2 = \\ 
    &2-\frac{2}{M} \mathrm{Re}\left(\sum_{j=1}^M \bra{\psi_j}U^\dagger C(\balpha) \ket{\psi_j} \right ).
\end{split} \end{equation}
To minimize the cost function, one can also maximize:
\begin{equation}
\mathrm{Re} \left( \sum_{j=1}^M \bra{\psi_j}U^\dagger C(\balpha) \ket{\psi_j} \right),
\end{equation}
such that for any specific gate $u_i$ in the circuit, can be written as $\mathrm{Re}(\tr(\cE u_i))$ (see also Fig.~\ref{fig:env_calc_tensor}). This is exactly what \qfac's inner step is maximizing~\cite[\S{III-A}]{kuklianskyQFactorDomainSpecificOptimizer2023a}, hence \qfacsasp uses the same updates to the gates as in \qfac and the update optimality proof~\cite[\S{III-A.1}]{kuklianskyQFactorDomainSpecificOptimizer2023a} holds here as well.

To lower the computation overhead of $\cE$ calculations, we compute only once $\{\bra{\psi_j}U^\dagger\}_{j=1}^M$ and denote it as the $A$ tensor. Moreover, when we compute $\{C(\balpha)\ket{\psi_j}\}_{j=1}^M$ we save the intermediate computation results in a list and denote it as $B$, see Fig.~\ref{fig:pre-comp} and Fig.~\ref{fig:env_calc_tensor}.

\begin{figure}[ht]
     \centering
     \begin{subfigure}[b]{0.2\linewidth}
         \centering
         \includegraphics[width=\linewidth]{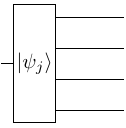}
         \caption{$B_0$}
         \label{fig:b0}
     \end{subfigure}
     \hfill
     \begin{subfigure}[b]{0.27\linewidth}
         \centering
         \includegraphics[width=\linewidth]{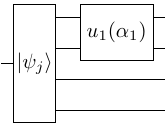}
         \caption{$B_1$}
         \label{fig:b1}
     \end{subfigure}
     \hfill
     \begin{subfigure}[b]{0.435\linewidth}
         \centering
         \includegraphics[width=\linewidth]{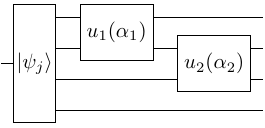}
         \caption{$B_2$}
         \label{fig:b2}
     \end{subfigure}
        \hfill
     \begin{subfigure}[b]{0.655\linewidth}
         \centering
         \includegraphics[width=\linewidth]{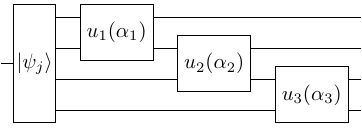}
         \caption{$B_3$}
         \label{fig:b3}
     \end{subfigure}
        \hfill
     \begin{subfigure}[b]{0.32\linewidth}
         \centering
         \includegraphics[width=\linewidth]{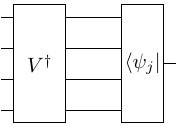}
         \caption{$A$}
         \label{fig:a0}
     \end{subfigure}
        \caption[Tensors pre-computation]{Tensors that are pre-computed and saved before each algorithm sweep. Recalculating the $A$ and $B_0$ tensors is done only when the training set $\{\ket{\psi_j}\}_{j=1}^M$ is updated. The pre-computation is a time-memory trade-off where we save intermediate computation results for later use and do not recompute them each time we calculate the environment matrix of a different gate. }
        \label{fig:pre-comp}
\end{figure}

\begin{figure*}[ht]
    \centering
    \includegraphics[width=\linewidth]{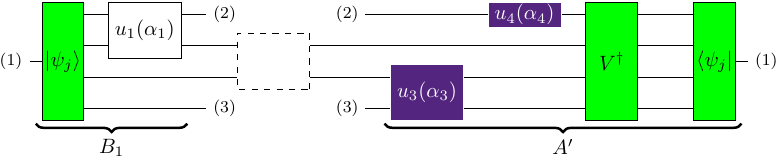}
    \caption[Environmental matrix calculation]{This figure depicts the environmental matrix calculation for the second gate in a four-gate circuit. Tensor legs with the same number will be traced together. Gate tensors colored in purple indicate parameters updated in the current sweep. Those in white denote gate tensors with parameters from the preceding sweep, while green represents tensors with consistent values across sweeps. We also note that for the tracing we use the previously calculated $B_1$ tensor, see Fig.~\ref{fig:pre-comp}.}
    \label{fig:env_calc_tensor}
\end{figure*}

The plateau detection mechanism in \qfacsasp checks if the training cost function has not sufficiently improved over $t$ consecutive iterations. In contrast, \qfacsp halts, indicating a plateau, if the cost function fails to improve enough in just a single iteration. This modification in \qfacsasp can improve the \acp{QoR} by sacrificing some runtime efficiency. Given that \qfacsasp is significantly faster than \qfac, this tradeoff is advantageous.

We implemented \qfacsasp on \acp{GPU} using JAX~\cite{jax2018github}, and our source code can be found at~\cite{BQSKitBqskitqfactorjax2023}. Some implementation details and a list of \qfacsa's hyperparameters are provided in Appendix~\ref{app:qfs_imp_d}. 

We think that the \textit{double-and-restart} approach is a good heuristic to use when the minimal number of training states $(M_{\text{min}})$ is unknown, as in the worse case we use $2M_{\text{min}}$ states, and the total amount of work is comparable to $4M_{\text{min}}$. An interesting question is how to choose the initial number of training states and the over-train threshold that controls the \emph{on-the-fly} generalization error detection. For circuits with more than a few qubits and dozens of gates, we observed that the first 2-3 attempts had the same runtime, although they had different number of training states. This is due to \ac{GPU} overheads, such as data transfer and code generation. Hence, starting with $M=2$ is probably suboptimal in real-life circumstances. 

Among all \qfacsa's hyperparameters, we would like to point out \textit{overtrain\_ratio} and \textit{number\_of\_training\_states}.  These hyperparameters determine the initial number of training states. If, during the instantiation, the normalized generalization error exceeds \textit{overtrain\_ratio}, the algorithm halts, doubles the number of training states, and restarts. The normalized generalization error is calculated by
\begin{equation}
    \frac{c_{val}}{c_{train}} - 1,
\end{equation}
where $c_{val}$ and $c_{train}$ represent the validation and training costs, respectively. The \textit{doubling-and-restart} process continues until the number of training states reaches $2^n$, where $n$ denotes the number of qubits.

We want to highlight the connection between \textit{overtrain\_ratio}, \textit{dist\_tol} (the unitary distance threshold parameter), and the resulting distance between the target unitary and instantiated circuit. When the instantiation is successful, the following two relations hold $c_{train}<d_{tol}$ and $\tfrac{c_{val}}{c_{train}} \le 1+otr$, where $d_{tol}$ and $otr$ represent \textit{dist\_tol}  and \textit{overtrain\_ratio} respectively. Then for a random state $\ket{\psi}$,
\begin{equation}
    ||U\ket{\psi} - C(\balpha)\ket{\psi}||^2 < d_{tol} (1+otr)
\end{equation}
holds with high probability. One might argue that minimizing $otr$ is preferable, yet doing so necessitates employing more training states. A simple remedy to mitigate generalization errors involves marginally reducing $d_{tol}$ while simultaneously increasing $otr$, ultimately leading to a faster instantiation, since we will need significantly fewer training states.

\section{Evaluation}\label{sec:qfs_evaluation}
We assess the performance of the instantiation algorithm using two metrics: success rate, which indicates the proportion of circuits from the benchmarks where the algorithm succeeded, and runtime, representing the total time taken by the algorithm to complete, regardless of the reason for termination. We used the same benchmarks and evaluation setup that was used in~\cite{kuklianskyQFactorDomainSpecificOptimizer2023a} to enable easy comparison.

The benchmarks used can be seen in Table.~\ref{tab:benchmarks}. They represent real circuits with 4--400 qubits, and varying depths of up to $\sim$200,000 gates. We performed the re-instantiate flow using 1727 random partitions from the above benchmarks and limited the runtime to ten minutes for partitions with fewer than nine qubits and two hours for the rest. In this flow, we take a partition of a circuit, calculate its unitary, and ask \qfacsasp to instantiate that unitary using the original partition circuit structure.

We compare \qfacsasp performance against the \ac{CPU} and \ac{GPU} versions of \qfacsp and a leading general-purpose numerical optimizer, which we will refer to as CERES~\cite{Agarwal_Ceres_Solver_2022}. In our comparison, we denote the Rust (\ac{CPU}) and Python+JAX (\ac{GPU}) implementations of \qfacsp by QF-R and QF-J respectively, and we denote \qfacsasp \ac{GPU} implementation as QFS-J.
c
We run \qfacsasp evaluation on NERSC's Perlmutter~\cite{perlmutter.arch} supercomputer. We used Perlmutter's hybrid GPU-CPU nodes. Each node has one AMD EPYC 7763 64-core processor, 256GB DDR4 DRAM, and four NVIDIA A100 \acp{GPU}, some have 40GB of RAM while others have 80GB. 

\qfacsa's hyperparameters used in the evaluation are: \textit{dist\_tol} $=10^{-10}$, \textit{diff\_tol\_r} $=10^{-3}$, \textit{plateau\_windows\_size}$=5$, $\beta=0$, \textit{number\_of\_training\_states}$=2$, \textit{min\_iter} $=6$, \textit{overtrain\_ratio}$=0.1$,   \textit{max\_iter} $=10^{6}$, and \textit{multistarts} $=32$ .

\begin{table}[htb]
\caption[Benchmarks and gate counts]{Benchmarks used and their gate counts, upper bound of $\sim$200,000.  The name suffix represents the number of qubits in the circuit, up to 400 qubits~\cite{kuklianskyQFactorDomainSpecificOptimizer2023a}.}
\tiny
\centering
\begin{tabular}{ |c| c| c||c| c| c| }
\hline
 Circuit & U3  & CNOT & Circuit & U3  & CNOT \\ 
 
 \hline
{\tt adder9 }& 64 & 98 & {\tt grover5 }& 80 & 48  \\
{\tt add17}& 348&232 & {\tt hhl8}& 3288 &2421\\
{\tt adder63 }& 2885 & 1405 & {\tt shor26 }& 20896 & 21072 \\

\hline
{\tt mult8}& 210&188 & {\tt hub4}& 155 & 180\\
{\tt mult16}& 1264&1128 & {\tt hub18 }& 1992 & 3541 \\

\hline
{\tt heis7}& 490&360 & {\tt tfim8}& 428&280 \\
{\tt heis8}& 570&420 & {\tt tfim16}& 916&600 \\
{\tt heis64}& 5050&3780 & {\tt tfim400 }& 88235 & 87670\\

\hline
{\tt qae11}& 176&110 & {\tt qpe8}& 519&372\\
{\tt qae13}& 247&156 & {\tt qpe10}& 1681&1260\\
{\tt qae33 }& 1617&1056 & {\tt qpe12}& 3582&2550 \\
{\tt qae81}& 7341 & 4840 &&&\\

\hline
{\tt qaoa5 }& 27 & 42 & {\tt vqe5}& 132&91\\
{\tt qaoa10 }& 40 & 85 & {\tt vqe12}& 4157&7640\\
{\tt qaoa12 }& 90 & 198 & {\tt vqe14}& 10792&20392\\
\hline

\end{tabular}
\label{tab:benchmarks}
\end{table}

\parah{Success rate} Fig.~\ref{fig:inst_succ_rate_qfs} holds a comparison of the success rates between the different instantiation algorithms and implementations. It is clear that QFS-J completely outperforms QF-J, while for the smaller partitions, QF-R and CERES outperform QFS-J. For partitions containing more than 8 qubits, QFS-J demonstrates the highest success rate, surpassing CERES, QF-J, and QF-R by factors of 41, 2.9, and 2.5 respectively. QFS-J employs more lenient criteria in its plateau detection mechanism; This, coupled with its reduced computational complexity, leads to better optimization performance compared to QF-J.

\begin{figure}[ht]
    \centering
    \includegraphics[width=\linewidth]{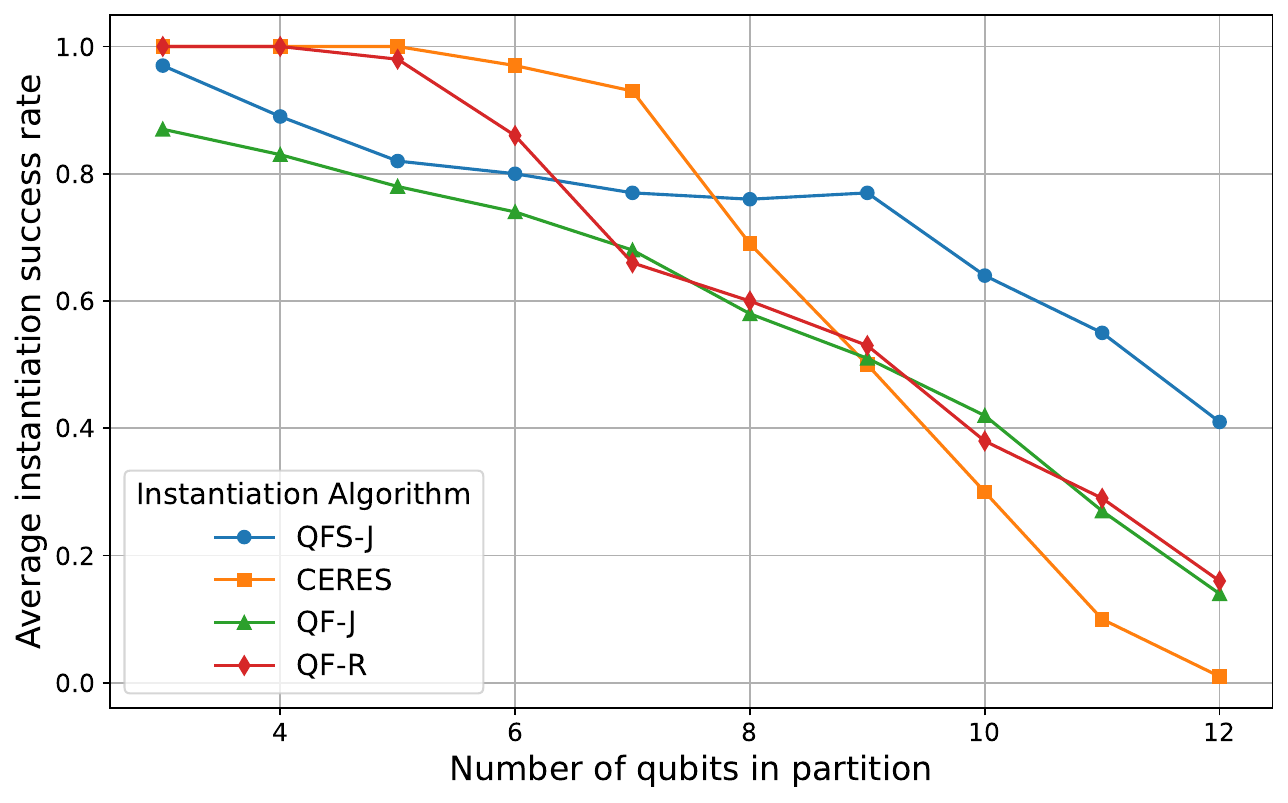}
    \caption{Instantiation Success Rate Comparison. The \ac{GPU} implementation of \qfacsasp significantly surpasses that of \qfac. Specifically, for circuits larger than 6 qubits, the \ac{GPU} version of \qfacsasp surpasses the \ac{CPU} version of \qfac. Furthermore, for circuits exceeding 7 qubits, the \ac{GPU} version of \qfacsasp outperforms the \ac{CPU} version of CERES.}
    \label{fig:inst_succ_rate_qfs}
\end{figure}

\parah{Execution speed} The relation between the runtime of QF-J and QFS-J can be seen in Fig.~\ref{fig:rel_inst_time_box_sworm}. Please see its caption for a detailed description of the swarm and box plots. We observe an overall average of 17.7X reduction in instantiation time and a 69X average reduction for partitions with more than 8 qubits.

\begin{figure}[ht]
    \centering
    \includegraphics[width=\linewidth]{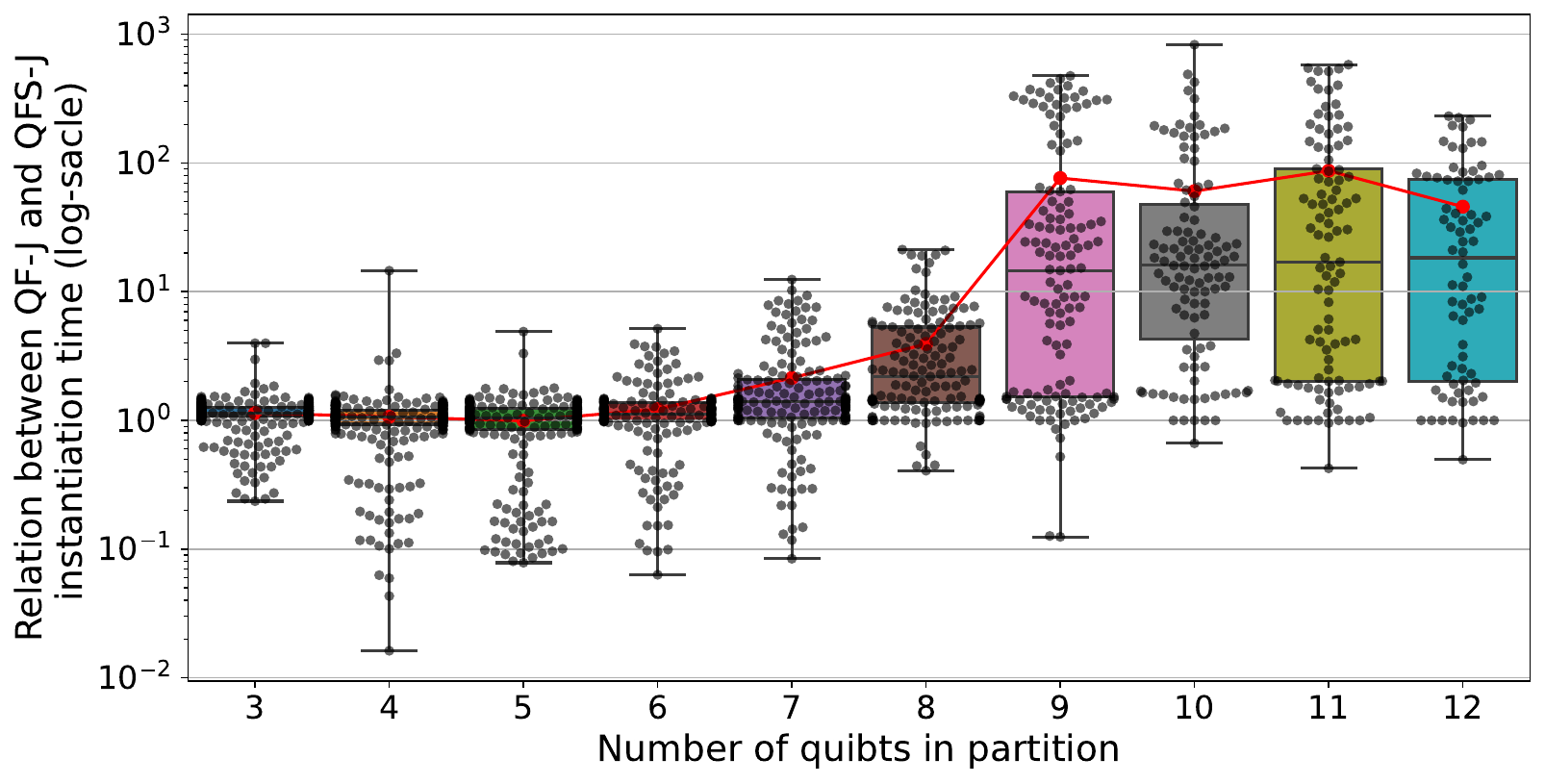}
    \caption[Instnatiation time comparison between \qfacsp and \qfacsa]{A swarm plot superimposed on a box plot of the relation between the instantiation time of \qfacsp and \qfacsasp plotted for different circuits size, shown on a log-scale. The whiskers extend to the maximum and the minimum values, while the box represents the interquartile range, which contains the middle 50\% of the data. A horizontal line inside the box represents the median. Each circle in the swarm plot represents a circuit, and it is plotted in a way that shows the distribution over the y-axis. The red markers represent the average runtime relation. From the plots, one can observe the significant improvement in the runtime of \qfacsasp compared to \qfac.}
    \label{fig:rel_inst_time_box_sworm}
\end{figure}

Fig.~\ref{fig:runtime_comp} illustrates a detailed comparative analysis of the instantiation performance between QFS-J and QF-J for partitions with 9 qubits, showing both the run time and the instantiation termination reason. We are providing the same type of graph for all of the partition sizes in Appendix~\ref{app:all_data}. Each point on the graphs represents an instantiation run, categorized by markers denoting success, failure to achieve the desired distance, and timeouts. Notably, markers are grouped based on the origin circuit of the partition, providing insights into performance trends across different circuit groups. On average, for partitions with 9 qubits, QFS-J is 76X times faster compared to QF-J.

\begin{figure*}[htb]
    \centering
    \includegraphics[width=1\linewidth]{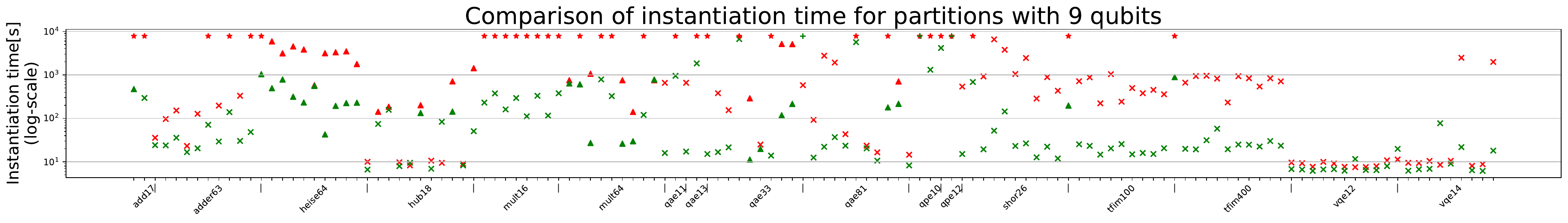}
    \caption{Instantiation runtime comparison between {\color{red} \qfac} (red) and \textcolor{green!50!black}{\qfacsa} (green) for partions with 9 qubits. Each mark on the graphs represents the runtime of a single instantiation. 'x' is a successful instantiation, '$\blacktriangle$' represents a run that finished, however, the desired distance was not achieved, while '*' and '+' are timeouts. The markers are grouped according to the partition's origin circuit, where the '|' marks the starting of a new circuit group. On average, \qfacsasp is 76X faster compared to \qfac and is able to find a good solution before timing out.}
    \label{fig:runtime_comp}
\end{figure*}

\parah{Circuit size} Fig.~\ref{fig:avg_inst_bin} shows the average success rate for the different instantiation algorithms binned according to the ratio $\frac{\#U3}{2^n}$, where and $n$ is the number of qubits and $\#U3$ is the number of U3 gates in the instantiated circuit. This type of binning was chosen because $\#U3$ is a relatively good proxy to the number of parameters the circuit has, and we know that \qfacsasp will need fewer training states as fewer parameters the circuit has. To further differentiate the circuit \qfacsasp is superior compared to other instantiation algorithms, we divide $\#U3$ with $2^n$, which reflects how compute-intensive the instantiation is going to be for \qfacsp and CERES. It is easy to see that the smaller the ratio is, the more \qfacsasp outperforms the other algorithms. 

\begin{figure}[tb]
    \centering
    \includegraphics[width=\linewidth]{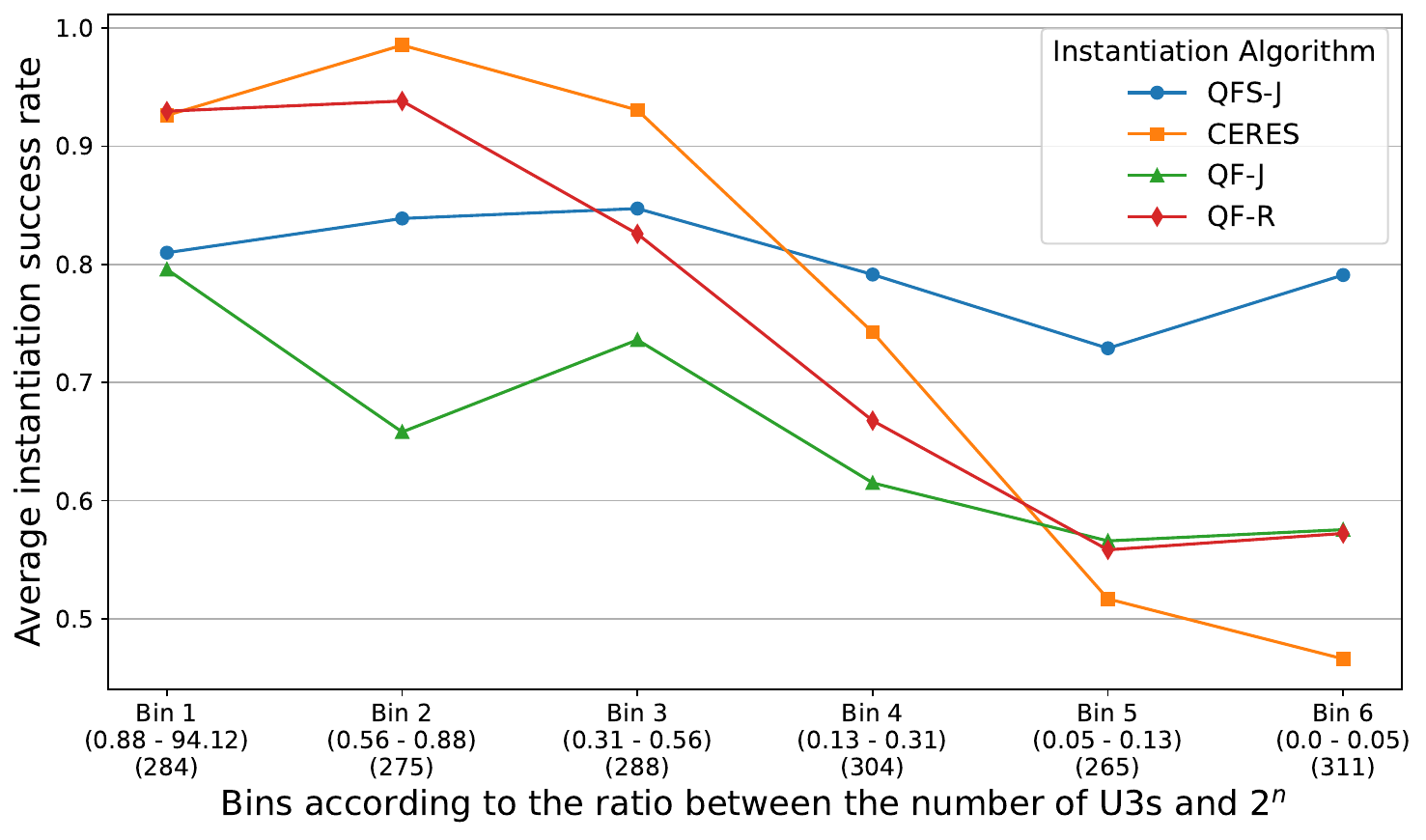}
    \caption[Average instantiation success rate]{Average instantiation success rate plotted for different instantiation algorithms and binned according to the ratio between the number of U3s and $2^n$, where $n$ is the number of qubits in the circuit. The x-axis label indicates the bin index, its boundaries, and the number of circuits the bin holds. As expected, QFS-J has a better success rate compared to the other algorithms when the above ratio is small, as the number of U3s the circuit has is an indicator for the number of parameters, and the fewer parameters it has the fewer training states QFS-J requires.}
    \label{fig:avg_inst_bin}
\end{figure}

\parah{Usage in synthesis} To conclude the evaluation, we incorporated \qfacsasp in BQSKit's~\cite{bqskit} re-synthesis gate deletion flow~\cite{qce22_ed}. This workflow involves initially partitioning the provided circuit, followed by a uni-directional sweep aimed at deleting one gate at a time while re-instantiating the reduced partition to its original unitary form. This divide-and-conquer approach effectively converts compilation into a task that can be executed in parallel, enhancing efficiency. Our focus lies in examining the scalability of large circuit compilations and assessing \qfacsa's effectiveness as a numerical optimizer, gauged by the runtime and the number of deleted gates within a circuit. We compared re-synthesis runtime and \acp{QoR} when using \qfacsasp against CERES and \qfac, utilizing the adder63, mult16, shor26, and qae11 circuits as benchmarks. This evaluation was conducted on eight hybrid nodes.

The results of our evaluation are presented in Table~\ref{tab:gate_del_flow_res}, showcasing the clear runtime superiority of \qfacsasp over \qfacsp and CERES. Notably, for the adder63, shor26, and qae11 circuits, we observe impressive speed-ups of 6X, 9X, and 4.5X  respectively. Furthermore, it is clear from all circuits that employing larger partitions during re-synthesis leads to better \ac{QoR}, albeit with the trade-off of longer runtime.

\begin{table*}[tb]
    \caption[Runtime and \acp{QoR} Comparison for Gate Deletion Flow]{Runtime and \acp{QoR} Comparison for Gate Deletion Flow. Each row represents a re-synthesis run on eight nodes with a 12-hour timeout. Noteworthy findings include a substantial 4-9X synthesis runtime improvement when using \qfacsasp over \qfac, and as expected, CERES is struggling on the big partitions and is more likely to timeout. Moreover, the utilization of larger partitions in synthesis leads to improved \ac{QoR}.}
    \centering
    \begin{tabular}{cccccl}
        Circuit & Instantiator & Partition Size& \#U3 & \#CNOT  & Runtime[s] \\
    \hline
         \multirow{4}{*}{\tt adder63} & QF-J & 7 &  1,724 & 1,250 & 31,413\\
         &  CERES & 7 & - & -  & timeout\\
         & QFS-J &  7& 1,685 & 1,245 & 5,109\\
         & QFS-J &  8 & 1,653 & 1,218 & 7,373\\
     \hline
         \multirow{3}{*}{\tt mult16} & QF-J & 6  & 573  & 874  & 24,660 \\
         & CERES & 6  & 569 & 874  & 21,463 \\
         &  QFS-J & 6  & 569  & 874  & 5,534 \\
         &  QFS-J & 9  & 488  & 836  & 20,839 \\
     \hline
         \multirow{4}{*}{\tt shor26} & QF-J  & 6  & 11,769  & 19,850  & 16,197 \\
         &  CERES & 6 & - & -  & timeout\\
         & QFS-J & 6  & 11,758  & 19,850  & 1,730 \\
         & QFS-J & 7  & 11,380  & 19,327  & 8,954 \\
    \hline
         \multirow{4}{*}{\tt qae11} & QF-J & 6 & 126 & 110 & 1,879 \\
         & CERES & 6 & 122 & 110 & 2,783 \\
         & QFS-J & 6 & 126 & 110 & 405 \\
         & QFS-J & 8 & 114 & 110 & 6,386
    \end{tabular}
    
    \label{tab:gate_del_flow_res}
\end{table*}

\section{Discussion}\label{sec:discussion}

Leveraging a bound on \ac{QML} generalization error, \qfacsasp significantly improves the runtime of quantum circuits instantiation. The source of the improvement comes from an exponential reduction in the computational complexity of the algorithm compared to other optimizers used for instantiation.

The instantiation runtime improvements that we report here, are somewhat skewed to the worse, as whenever an instantiation timed out we registered it as 10 or  720 minutes, depending on the circuit size. Hence, if we were able to run over the time limit, then the runtime gap between \qfacsp and \qfacsasp would increase substantially.

The enhancement in \qfacsa's instantiation success rate stems from its increased speed. This not only enables the algorithm to complete before the timeout but also permits the application of a more forgiving policy for plateau detection. Consequently, \qfacsasp can uncover more optimal solutions on challenging optimization planes. 

When evaluating \qfacsa's impact on circuit re-synthesis, we observe a direct decrease in overall runtime. Furthermore, leveraging a more scalable instantiation algorithm enables the utilization of larger partitions, leading to the discovery of additional optimization opportunities and resulting in a reduction in circuit size. During our benchmarks, we noticed that for some circuits the partitioner we used was able to only cover part of the circuit with large partitions, impeding optimization opportunities. Appendix~\ref{app:parts_stats} provides the coverage statistics. Further investigation into the partitioning algorithm, although beyond this paper's scope, is a promising area for future research.

Our benchmarking results analysis can serve as a guide for quantum compilers to choose the appropriate instantiation algorithm according to simple metrics of the instantiated circuit. For example, if there are fewer than five qubits, it is better to use \qfac-RUST or CERES, and for the larger qubit count, one should look at the ratio between the number of parameters the circuit has and compare it to $2^n$ to decide whether to use \qfacsasp or \qfac. No single optimizer is perfect for all partitions, but with some hyperparameter tweaking and a decision regarding which instantiator to use on each partition, the overall performance of the compiler can be improved.

In this work, we assessed the performance of \qfacsasp on qubit gates. However, it is worth noting that the same algorithm is applicable to qudit gates, where the speedup compared to \qfacsp would be even more significant, reducing the complexity from $O(d^{2n})$ to $O(Md^{n})$.

An interesting follow-up research would be to perform unitary instantiation using one of the \ac{QML} frameworks such as TensorFlow Quantum or torchquantum. In this paper, we have shown a reduction between instantiation to a traditional \ac{QML} flow. Given the considerable engineering efforts invested in these frameworks, it becomes even more compelling to compare their runtime and \ac{QoR} with \qfacsasp.

\section*{Acknowledgements}

The research presented in this paper (LC) was supported by the Laboratory Directed Research and Development (LDRD) program of Los Alamos National Laboratory (LANL) under project numbers 20230049DR and 20230067DR. CI was supported by the U.S. DOE under contract DE5AC02-05CH11231, through the Office of Advanced Scientific Computing Research (ASCR), under the Accelerated Research in Quantum Computing (ARQC) program.

This research used resources of the National Energy Research Scientific Computing Center (NERSC), a U.S. Department of Energy Office of Science User Facility located at Lawrence Berkeley National Laboratory, operated under Contract No. DE-AC02-05CH11231 using NERSC award DDR-ERCAPm4141.

\bibliographystyle{quantum}
\bibliography{qfactor_sample}

\begin{thebibliography}{10}

\bibitem{vqe}
Jarrod~R McClean, Jonathan Romero, Ryan Babbush, and Alán Aspuru-Guzik.
\newblock ``The theory of variational hybrid quantum-classical algorithms''.
\newblock \href{https://dx.doi.org/10.1088/1367-2630/18/2/023023}{New Journal of Physics {\bf 18}, 023023}~(2016).

\bibitem{qaoa}
Edward Farhi, Jeffrey Goldstone, and Sam Gutmann.
\newblock ``A {{Quantum Approximate Optimization Algorithm}}''~(2014).
\newblock  \href{http://arxiv.org/abs/1411.4028}{arXiv:1411.4028}.

\bibitem{cpflow}
Nikita~A. Nemkov, Evgeniy~O. Kiktenko, Ilia~A. Luchnikov, and Aleksey~K. Fedorov.
\newblock ``Efficient variational synthesis of quantum circuits with coherent multi-start optimization''.
\newblock \href{https://dx.doi.org/10.22331/q-2023-05-04-993}{Quantum {\bf 7}, 993}~(2023).

\bibitem{squander1}
Péter Rakyta and Zoltán Zimborás.
\newblock ``Approaching the theoretical limit in quantum gate decomposition''~(2021).
\newblock  \href{http://arxiv.org/abs/quant-ph/2109.06770}{arXiv:quant-ph/2109.06770}.

\bibitem{nacl}
Lukasz Cincio, Kenneth Rudinger, Mohan Sarovar, and Patrick~J. Coles.
\newblock ``Machine learning of noise-resilient quantum circuits''.
\newblock \href{https://dx.doi.org/10.1103/PRXQuantum.2.010324}{PRX Quantum {\bf 2}, 010324}~(2021).

\bibitem{bqskit}
Ed~Younis, Costin~C Iancu, Wim Lavrijsen, Marc Davis, Ethan Smith, et~al.
\newblock ``Berkeley quantum synthesis toolkit (bqskit) v1''~(2021).

\bibitem{qfast}
E.~Younis, K.~Sen, K.~Yelick, and C.~Iancu.
\newblock ``Qfast: Conflating search and numerical optimization for scalable quantum circuit synthesis''.
\newblock In 2021 IEEE International Conference on Quantum Computing and Engineering (QCE).
\newblock \href{https://dx.doi.org/10.1109/QCE52317.2021.00041}{Pages 232--243}.
\newblock IEEE Computer Society~(2021).

\bibitem{qsearch}
Marc~G. Davis, Ethan Smith, Ana Tudor, Koushik Sen, Irfan Siddiqi, and Costin Iancu.
\newblock ``Towards optimal topology aware quantum circuit synthesis''.
\newblock In 2020 IEEE International Conference on Quantum Computing and Engineering (QCE).
\newblock \href{https://dx.doi.org/10.1109/QCE49297.2020.00036}{Pages 223--234}.
\newblock ~(2020).

\bibitem{qce22_ed}
E.~Younis and C.~Iancu.
\newblock ``Quantum circuit optimization and transpilation via parameterized circuit instantiation''.
\newblock In 2022 IEEE International Conference on Quantum Computing and Engineering (QCE).
\newblock \href{https://dx.doi.org/10.1109/QCE53715.2022.00068}{Pages 465--475}.
\newblock ~(2022).

\bibitem{cerezoVariationalQuantumAlgorithms2021}
M.~Cerezo, Andrew Arrasmith, Ryan Babbush, Simon~C. Benjamin, Suguru Endo, Keisuke Fujii, Jarrod~R. McClean, Kosuke Mitarai, Xiao Yuan, Lukasz Cincio, and Patrick~J. Coles.
\newblock ``Variational quantum algorithms''.
\newblock \href{https://dx.doi.org/10.1038/s42254-021-00348-9}{Nature Reviews Physics {\bf 3}, 625--644}~(2021).

\bibitem{benedettiParameterizedQuantumCircuits2019}
Marcello Benedetti, Erika Lloyd, Stefan Sack, and Mattia Fiorentini.
\newblock ``Parameterized quantum circuits as machine learning models''.
\newblock \href{https://dx.doi.org/10.1088/2058-9565/ab4eb5}{Quantum Science and Technology {\bf 4}, 043001}~(2019).

\bibitem{biamonteQuantumMachineLearning2017}
Jacob Biamonte, Peter Wittek, Nicola Pancotti, Patrick Rebentrost, Nathan Wiebe, and Seth Lloyd.
\newblock ``Quantum machine learning''.
\newblock \href{https://dx.doi.org/10.1038/nature23474}{Nature {\bf 549}, 195--202}~(2017).

\bibitem{beerTrainingDeepQuantum2020}
Kerstin Beer, Dmytro Bondarenko, Terry Farrelly, Tobias~J. Osborne, Robert Salzmann, Daniel Scheiermann, and Ramona Wolf.
\newblock ``Training deep quantum neural networks''.
\newblock \href{https://dx.doi.org/10.1038/s41467-020-14454-2}{Nature Communications {\bf 11}, 808}~(2020).

\bibitem{kuklianskyQFactorDomainSpecificOptimizer2023a}
Alon Kukliansky, Ed~Younis, Lukasz Cincio, and Costin Iancu.
\newblock ``{{QFactor}}: {{A Domain-Specific Optimizer}} for {{Quantum Circuit Instantiation}}''.
\newblock In 2023 {{IEEE International Conference}} on {{Quantum Computing}} and {{Engineering}} ({{QCE}}).
\newblock \href{https://dx.doi.org/10.1109/QCE57702.2023.00096}{Volume~01, pages 814--824}.
\newblock ~(2023).

\bibitem{abbasPowerQuantumNeural2021}
Amira Abbas, David Sutter, Christa Zoufal, Aurelien Lucchi, Alessio Figalli, and Stefan Woerner.
\newblock ``The {{Power}} of {{Quantum Neural Networks}}''.
\newblock \href{https://dx.doi.org/10.1038/s43588-021-00084-1}{Nature Computational Science {\bf 1}, 403--409}~(2021).

\bibitem{caroGeneralizationQuantumMachine2022}
Matthias~C. Caro, Hsin-Yuan Huang, M.~Cerezo, Kunal Sharma, Andrew Sornborger, Lukasz Cincio, and Patrick~J. Coles.
\newblock ``Generalization in quantum machine learning from few training data''.
\newblock \href{https://dx.doi.org/10.1038/s41467-022-32550-3}{Nature Communications {\bf 13}, 4919}~(2022).

\bibitem{banchiGeneralizationQuantumMachine2021}
Leonardo Banchi, Jason Pereira, and Stefano Pirandola.
\newblock ``Generalization in {{Quantum Machine Learning}}: {{A Quantum Information Standpoint}}''.
\newblock \href{https://dx.doi.org/10.1103/PRXQuantum.2.040321}{PRX Quantum {\bf 2}, 040321}~(2021).

\bibitem{duEfficientMeasureExpressivity2022}
Yuxuan Du, Zhuozhuo Tu, Xiao Yuan, and Dacheng Tao.
\newblock ``Efficient {{Measure}} for the {{Expressivity}} of {{Variational Quantum Algorithms}}''.
\newblock \href{https://dx.doi.org/10.1103/PhysRevLett.128.080506}{Physical Review Letters {\bf 128}, 080506}~(2022).

\bibitem{orus2014practical}
Román Orús.
\newblock ``A practical introduction to tensor networks: Matrix product states and projected entangled pair states''.
\newblock \href{https://dx.doi.org/https://doi.org/10.1016/j.aop.2014.06.013}{Annals of Physics {\bf 349}, 117--158}~(2014).

\bibitem{nocedal1980updating}
Jorge Nocedal.
\newblock ``Updating quasi-newton matrices with limited storage''.
\newblock \href{https://dx.doi.org/https://doi.org/10.2307/2006193}{Mathematics of computation {\bf 35}, 773--782}~(1980).

\bibitem{liu1989limited}
Dong~C Liu and Jorge Nocedal.
\newblock ``On the limited memory bfgs method for large scale optimization''.
\newblock \href{https://dx.doi.org/https://doi.org/10.1007/BF01589116}{Mathematical programming {\bf 45}, 503--528}~(1989).

\bibitem{ranganathan2004levenberg}
Ananth Ranganathan.
\newblock ``The levenberg-marquardt algorithm''.
\newblock \url{ https://sites.cs.ucsb.edu/~yfwang/courses/cs290i_mvg/pdf/LMA.pdf}~(2004).

\bibitem{aurelien2019hands}
Aur{\'e}lien G{\'e}ron.
\newblock ``{Hands-on Machine Learning with scikit-learn, Keras, \& TensorFlow}''.
\newblock O'Reilly Media. Sebastopol, CA, USA~(2019).
\newblock 2nd edition.

\bibitem{jax2018github}
James Bradbury, Roy Frostig, Peter Hawkins, Matthew~James Johnson, Chris Leary, Dougal Maclaurin, George Necula, Adam Paszke, Jake Vander{P}las, Skye Wanderman-{M}ilne, and Qiao Zhang.
\newblock ``{JAX}: composable transformations of {P}ython+{N}um{P}y programs''.
\newblock \url{http://github.com/google/jax}.

\bibitem{BQSKitBqskitqfactorjax2023}
A.~Kukliansky.
\newblock ``{{GPU implementation of QFactor circuit instantiation using JAX}}''.
\newblock \url{https://github.com/BQSKit/bqskit-qfactor-jax/}.

\bibitem{Agarwal_Ceres_Solver_2022}
Sameer Agarwal, Keir Mierle, and The Ceres~Solver Team.
\newblock ``{Ceres Solver}''.
\newblock \url{https://github.com/ceres-solver/ceres-solver}~(2022).

\bibitem{perlmutter.arch}
``Perlmutter architecture''.
\newblock \url{https://docs.nersc.gov/systems/perlmutter/architecture/}~(2023).

\end{thebibliography}

\onecolumn \newpage
\appendix

\section{\qfacsasp Implementation Details}\label{app:qfs_imp_d}
In this appendix, we provide more details about \qfacsasp implementation.
We list all of \qfacsa's hyperparameters and provide some best practices when using our implementation.

We generated the random input test states by extracting columns of a random unitary matrix, which was generated using the {\tt unitary\_group} functionality in SciPy. Doubling the size of the training state does not involve randomizing the unitaries of the gates, as we have found that this does not impact the convergence speed. Instead, it only consumes time by generating numerous random unitaries for each gate.

During our testing of the implementation, we encountered \ac{GPU} \ac{OOM} exceptions. We did not find a way to predict our \ac{GPU} memory usage in JAX; hence we implemented a recursive trial-and-error when running \qfacsa. If while running \qfacsasp we catch a \ac{GPU} \ac{OOM} exception, we would iteratively run two runs of \qfacsasp but with half the number of multistarts. We also want to mention that JAX has an environment variable, {\footnotesize XLA\_PYTHON\_CLIENT\_ALLOCATOR}, that forces the runtime to always release memory buffers to the \ac{GPU} when they are no longer needed, avoiding \ac{OOM} due to fragmentation with the potential runtime cost of reallocating memory.

Following is a list of \qfacsa's hyperparameters. They control the termination conditions of the algorithm, generalization error threshold, gate update policy, and randomization:
\begin{itemize}
    \item \textit{dist\_tol}: When the average distance between the training output states and those generated by the circuit~\eqref{eq:qfactor_sample_cost_function} is lower than \textit{dist\_tol}, the algorithm stops.
    
    \item \textit{plateau\_windows\_size} and \textit{diff\_tol\_r}: Control the plateau-detection mechanism. The algorithm will terminate due to plateauing when on consecutive \textit{plateau\_windows\_size} iterations the cost function did not relatively improve by \textit{diff\_tol\_r}. The algorithm checks if the relative improvement has been met by calculating the following relation:
    \begin{equation}
       |c_{i-1}| - |c_{i}| >  \textit{diff\_tol\_r} * |c_{i}|,
    \end{equation}
    where $c_{i}$ is the cost function value after iteration $i$.

    \item \textit{number\_of\_training\_states} and \textit{overtrain\_ratio}: These set the initial number of training states. If during the instantiation the algorithm detects that the normalized generalization error is bigger than \textit{overtrain\_ratio}, it stops, doubles the number of training states, and restarts. The normalized generalization error is computed by:
    \begin{equation}
        \frac{c_{val}}{c_{train}} - 1,
    \end{equation}
    where $c_{val}$ and $c_{train}$ are the validation and training costs, respectively. The \textit{double-and-restart} stops when the number of training states reaches $2^n$, where $n$ is the number of qubits.
    
    \item \textit{min\_iter}: Sets the minimum number of iterations for \qfacsasp to complete before stopping due to plateauing or overtraining.
    
    \item \textit{max\_iter}: Sets the maximum number of iterations for \qfacsa. The algorithm will always stop when it reaches this limit.
    
    \item \textit{multistarts} and \textit{seed}: To overcome the local minimum problem, one can run \qfacsasp with various initial gate unitaries, in the hope that at least one of the runs will converge to a good solution. The initial random unitaries for each gate are controlled by a \textit{seed} parameter.

    \item \textit{Beta}: Serves as a regularization parameter that governs the retention of the previous value during the gate update step. Instead of conducting the \ac{SVD} operation solely on the environment $\cE$, it is performed on the linear combination:

    \begin{equation}
        (1-\beta)\cE + \beta * u^\dagger.
    \end{equation}

    This parameter proves beneficial in addressing slow convergence issues encountered in circuits with inter-gate dependencies, where local optimization methods may fall short. When $\beta = 0$, the update becomes \textit{full} as described previously, whereas $\beta = 1$ results in no update taking place.
\end{itemize}

From our experience, the following are some good initial hyperparameter values: \textit{dist\_tol} $=10^{-8}$, \textit{plateau\_windows\_size}$=5$, \textit{diff\_tol\_r} $=10^{-5}$, \textit{number\_of\_training\_states}$=2$, \textit{overtrain\_ratio}$=0.1$,  \textit{min\_iter} $=6$, \textit{max\_iter} $=10^{4}$, \textit{multistarts} $=16$, and $\beta=0$.
By modifying the above-mentioned hyperparameters, one can easily adjust the tradeoff between result quality and execution time. If one wishes to get results faster, decrease \textit{max\_iter}, increase \textit{diff\_tol\_r} and \textit{diff\_tol\_a}. Alternately, one might increase $\beta$ or \textit{multistarts} to find better results with additional execution overhead.

\section{Instantiation Performance Comparison Between \qfacsasp and \qfacsp for Circuits with 3 to 12 Qubits }\label{app:all_data}

In this appendix, we provide the complete instantiation performance comparison for circuits with 3 to 12 qubits using \qfacsasp and \qfac. This is an extension to data presented in Fig.~\ref{fig:runtime_comp}.

A notable observation from the analysis is the widening performance gap between \qfacsasp and \qfacsp as the circuit size increases with more qubits. This is expected due to the differences in computational complexity. As the partition size grows, \qfacsasp experiences significantly fewer timeouts compared to \qfacsp, enabling it to either reach a good solution or stop upon detecting a plateau. In the graph, we sorted the instantiations based on the origin circuit of each partition and used different markers to indicate the termination type. This visualization helps illustrate why \qfacsasp has a higher success rate compared to \qfacsp for partitions with 3 qubits. The key difference lies in the plateau detection mechanism: \qfacsasp employs a more lenient approach, enabling it to find good solutions for all of the Heisenberg partitions (as shown on the right side of the graph), whereas \qfacsp gives up on those.

\begin{figure}[ht]
    \centering
    \includegraphics[width=\linewidth]{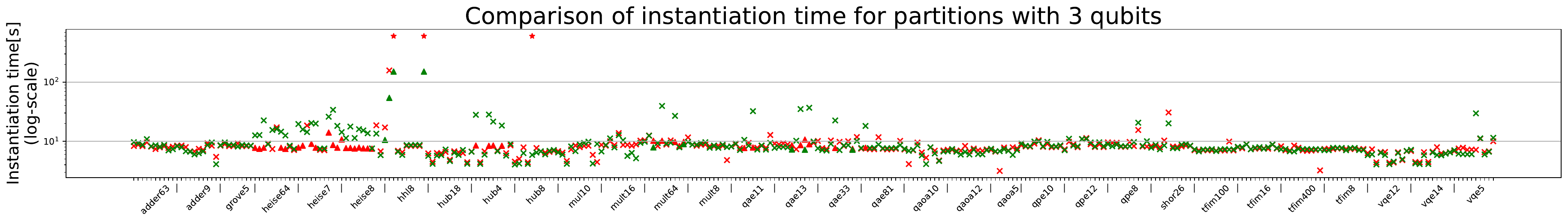}
    \includegraphics[width=\linewidth]{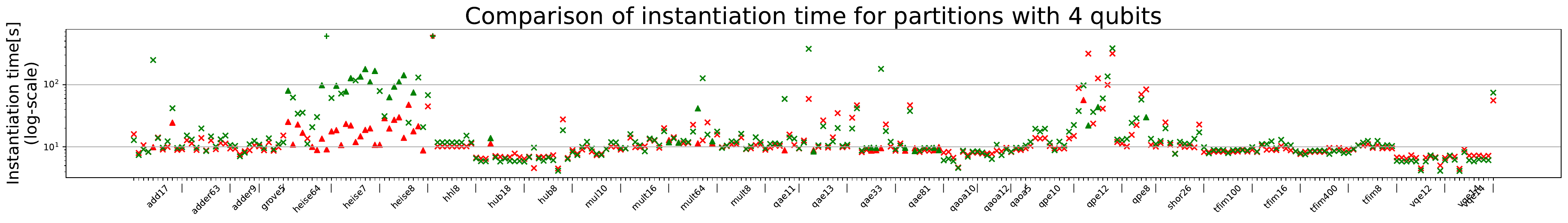}
    
 \caption{Instantiation runtime comparison between {\color{red} \qfac} (red) and \textcolor{green!50!black}{\qfacsa} (green). Each graph represents a different partition size varying from three to twelve. Each mark on the graphs represents the runtime of a single instantiation. 'x' is a successful instantiation, '$\blacktriangle$' represents a run that finished, however, the desired distance was not achieved, while '*' and '+' are timeouts. The markers are grouped according to the partition's origin circuit, where the '|' marks the starting of a new circuit group. We can observe that as the circuit has more qubits the performance gap between \qfacsasp and \qfacsp increases up to 100X, ultimately allowing \qfacsasp to instantiate bigger circuits.}
    
\end{figure}
\begin{figure}[htbp!]\ContinuedFloat
    \centering
    \includegraphics[width=\linewidth]{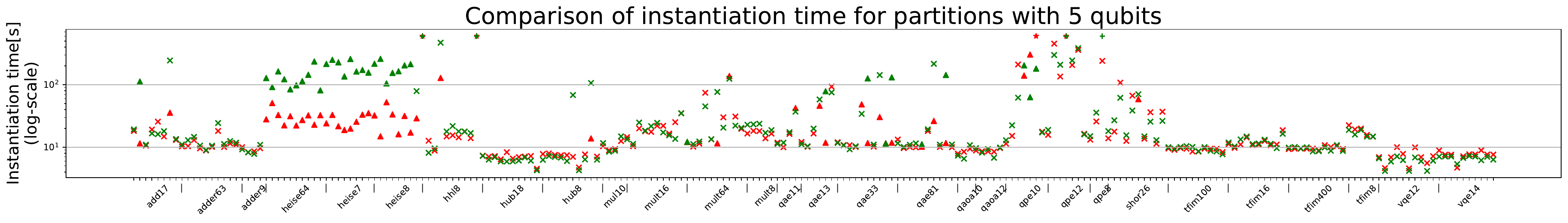}
    \includegraphics[width=\linewidth]{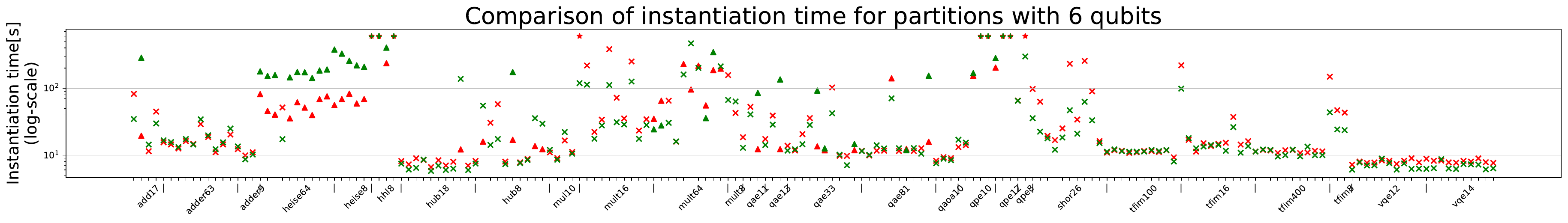}
    \includegraphics[width=\linewidth]{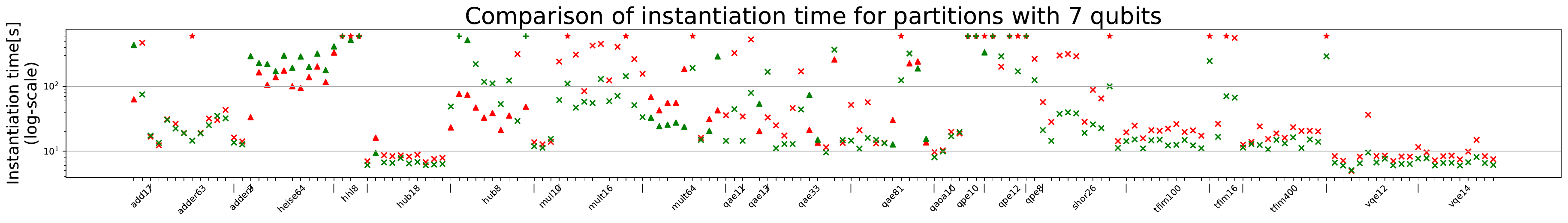}
    \includegraphics[width=\linewidth]{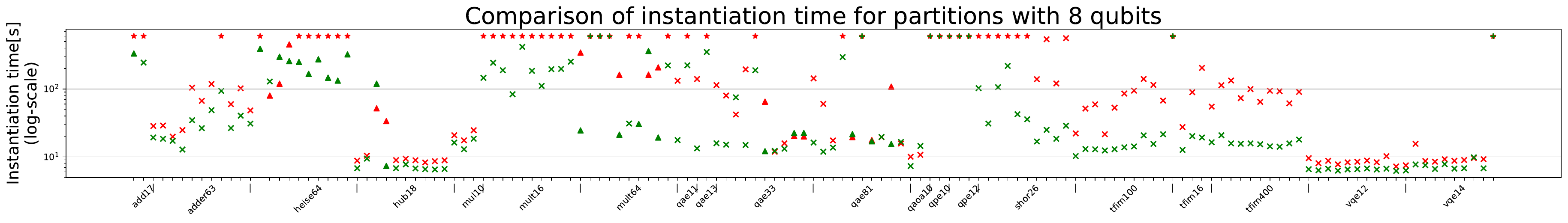}
    \includegraphics[width=\linewidth]{figures/inst_comp_9q.pdf}
    \includegraphics[width=\linewidth]{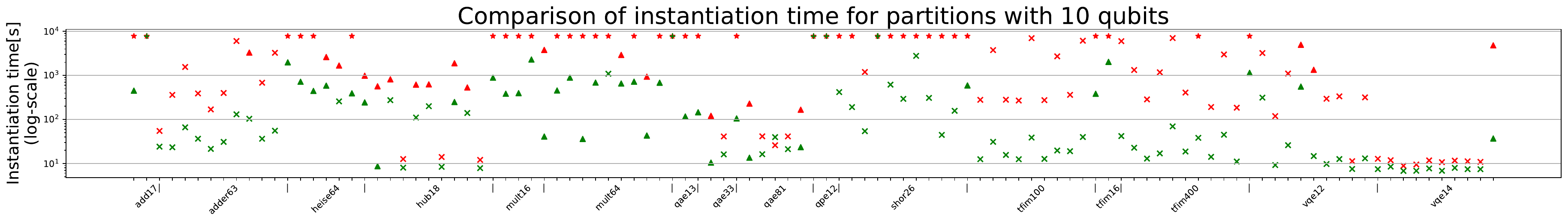}
    \includegraphics[width=\linewidth]{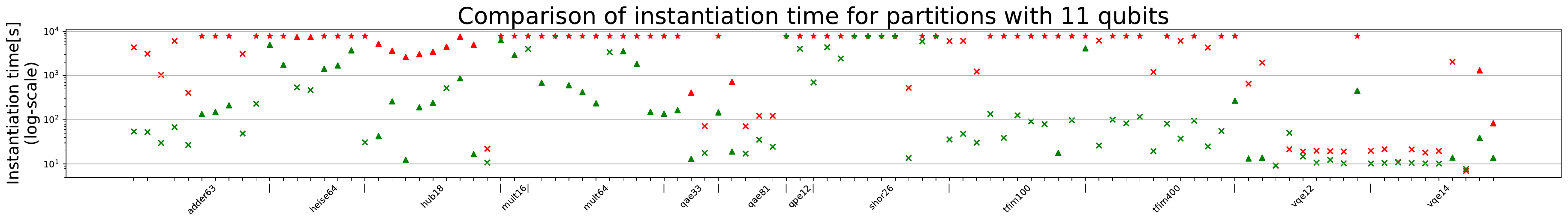}
    \includegraphics[width=\linewidth]{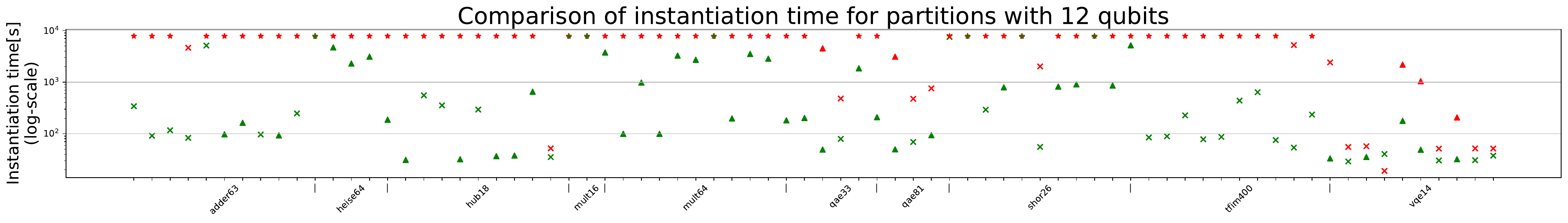}
    \caption{See caption on the previous page.}
   
\end{figure}

\section{Partitioner Coverage Statistics}\label{app:parts_stats}

During the evaluation of \qfacsasp as part of a re-synthesis flow, we observed that for some circuits BQSkit's "QuickPartitioner" allocates a significant portion of the circuit area to partitions with a small number of qubits, thus impeding optimization opportunities. In this appendix, we present BQSkit's "QuickPartitioner" performance on different circuits.

\begin{figure}[htbp!]
    \centering
    \includegraphics[width=0.445\linewidth]{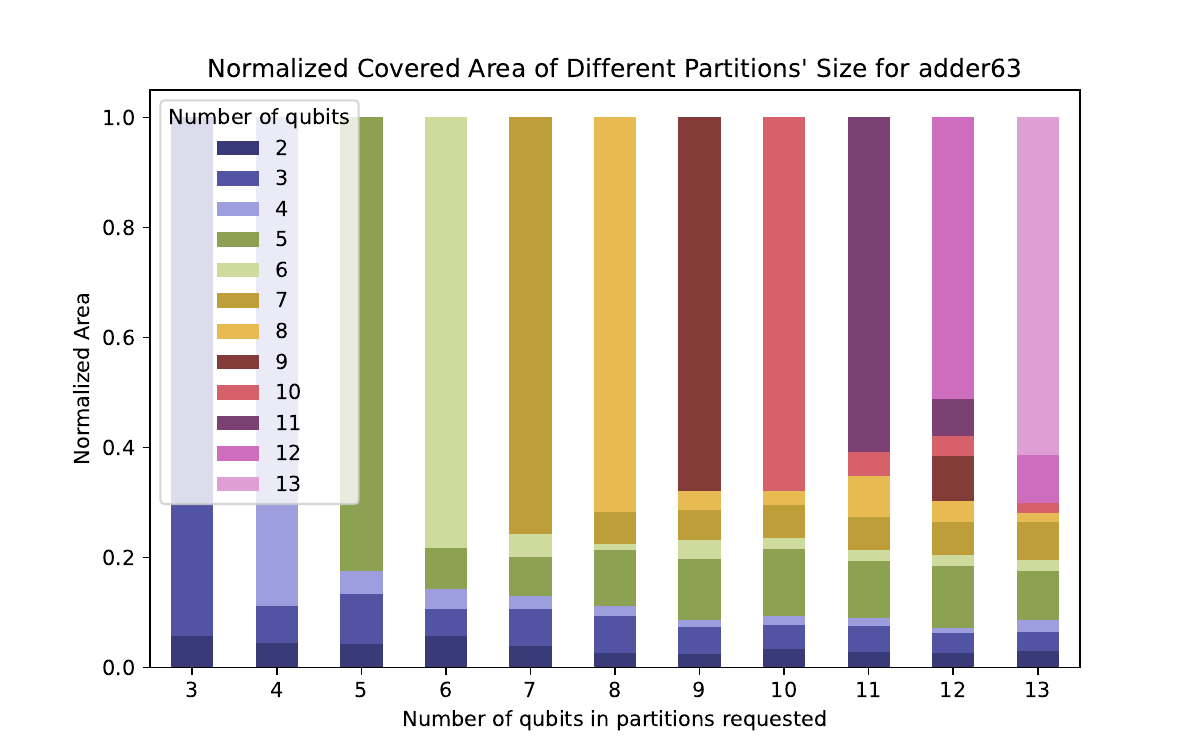}
    \includegraphics[width=0.445\linewidth]{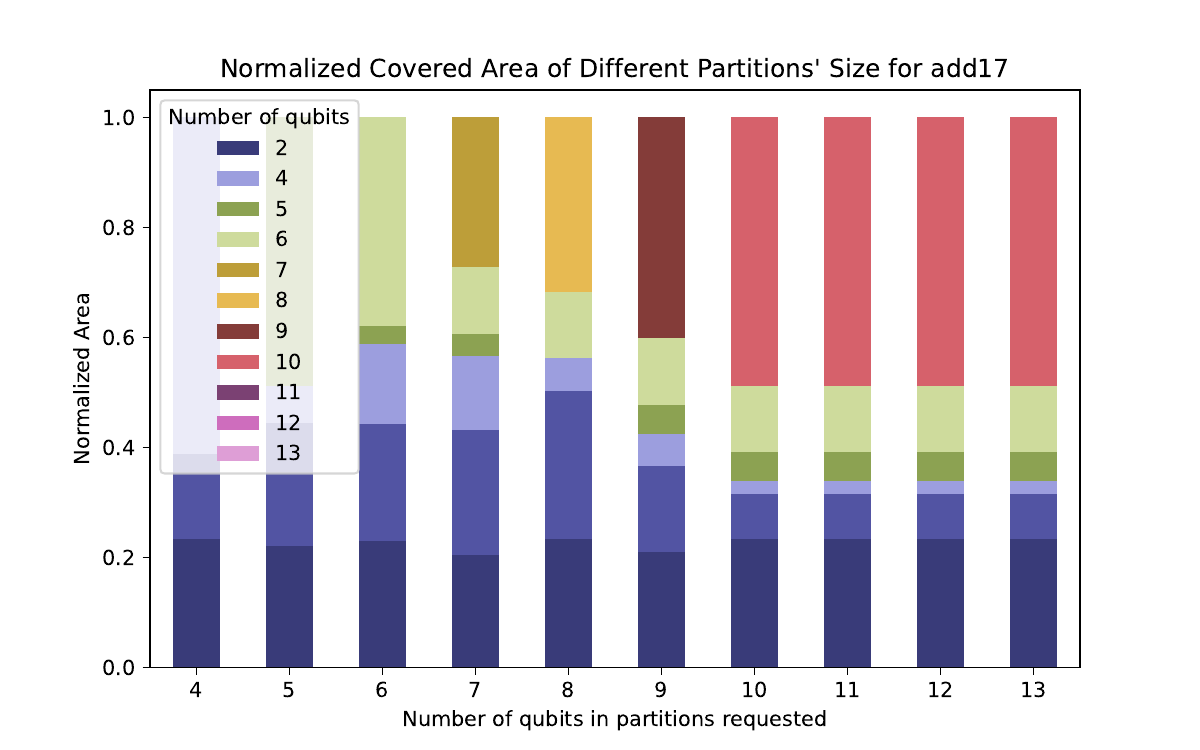}
    \includegraphics[width=0.445\linewidth]{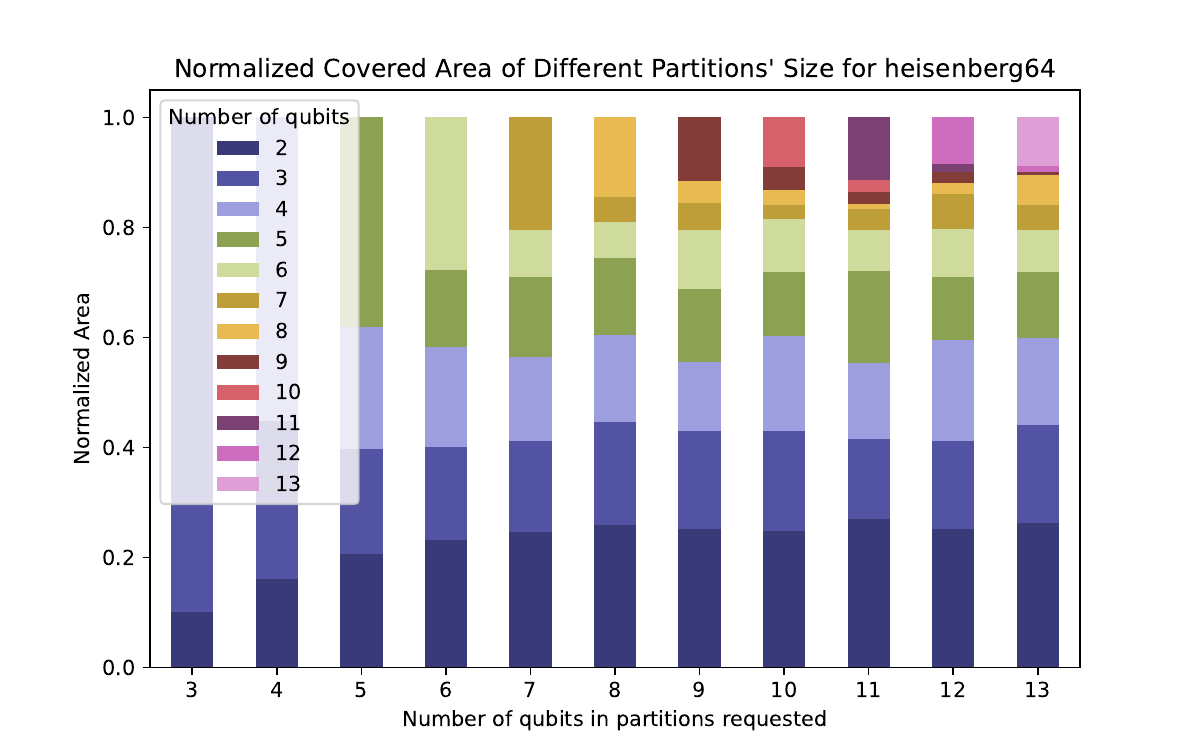}
    \includegraphics[width=0.445\linewidth]{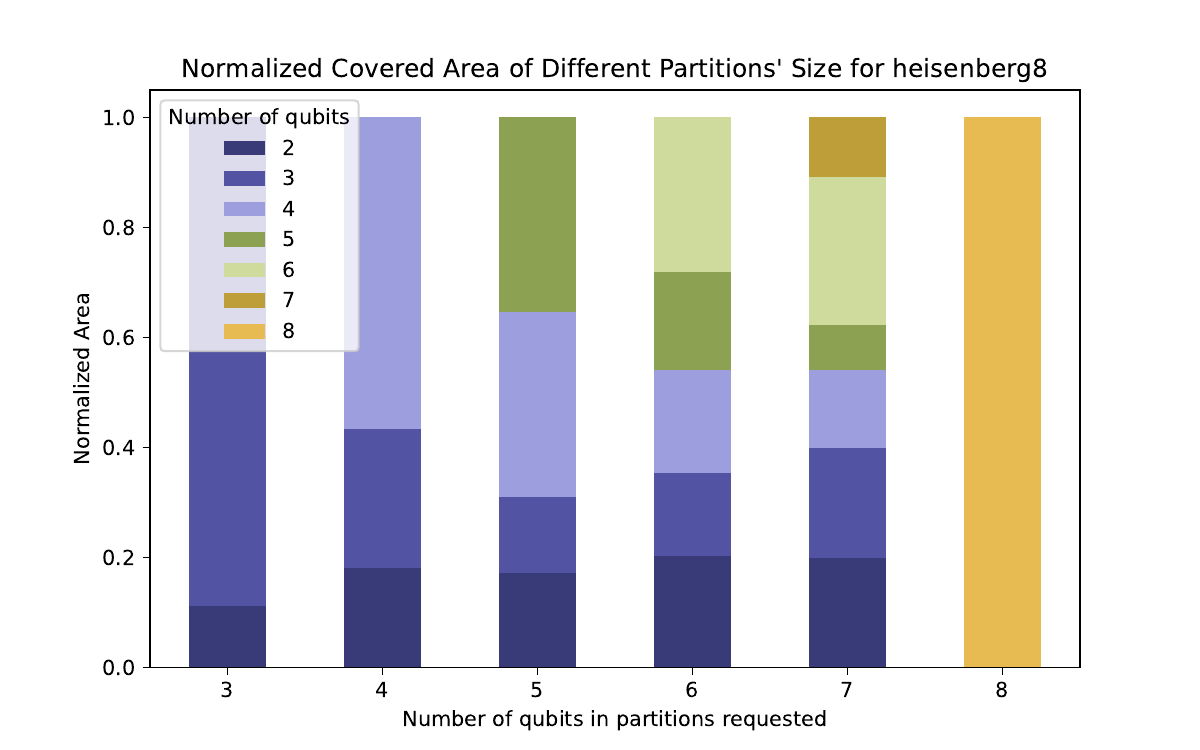}   
    \includegraphics[width=0.445\linewidth]{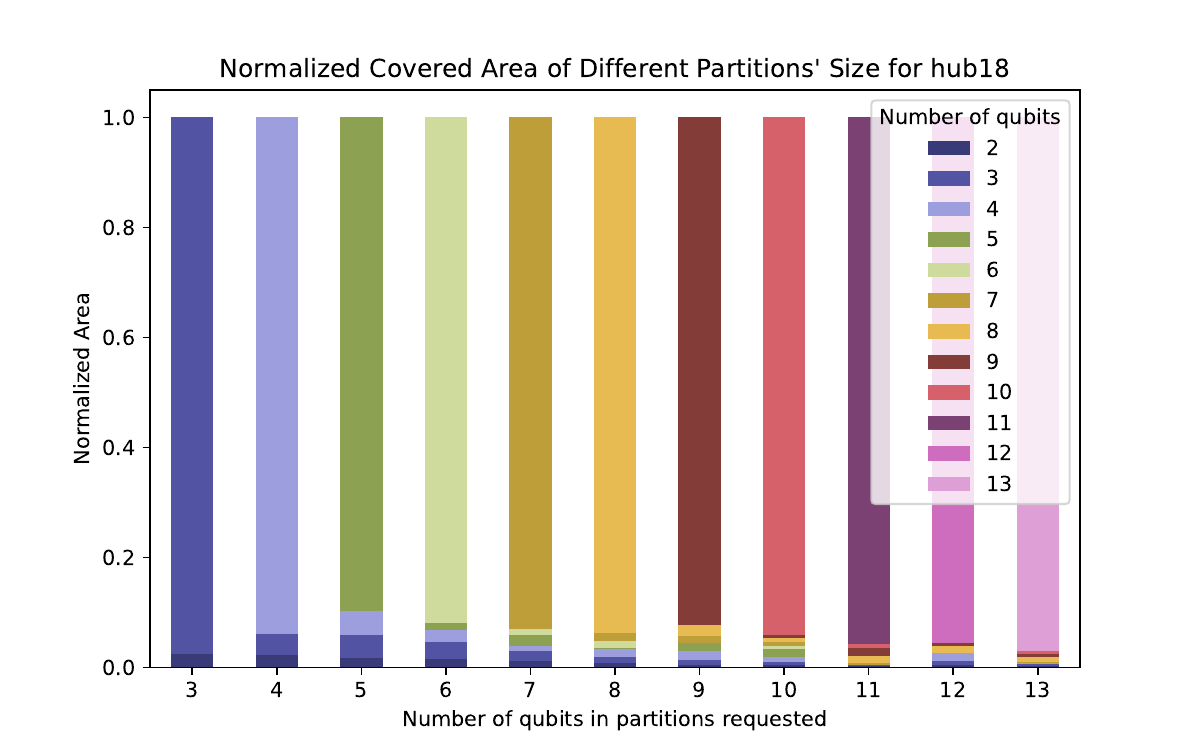}
    \includegraphics[width=0.445\linewidth]{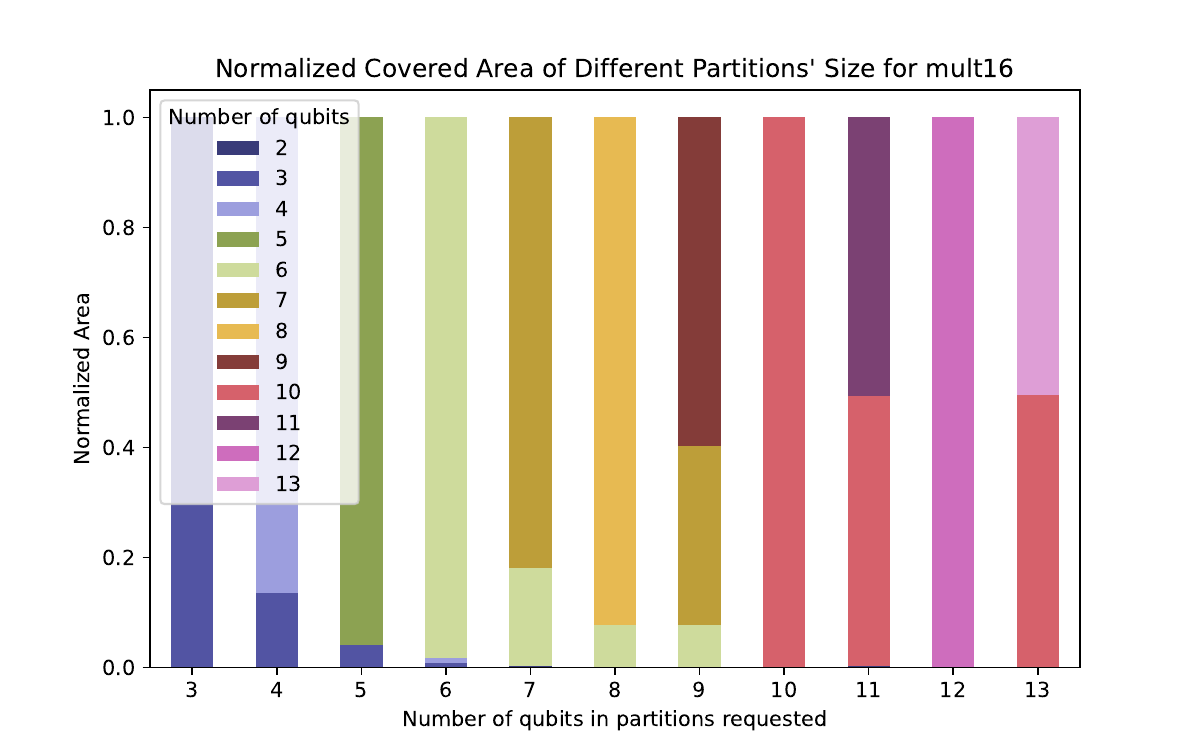}
    \includegraphics[width=0.445\linewidth]{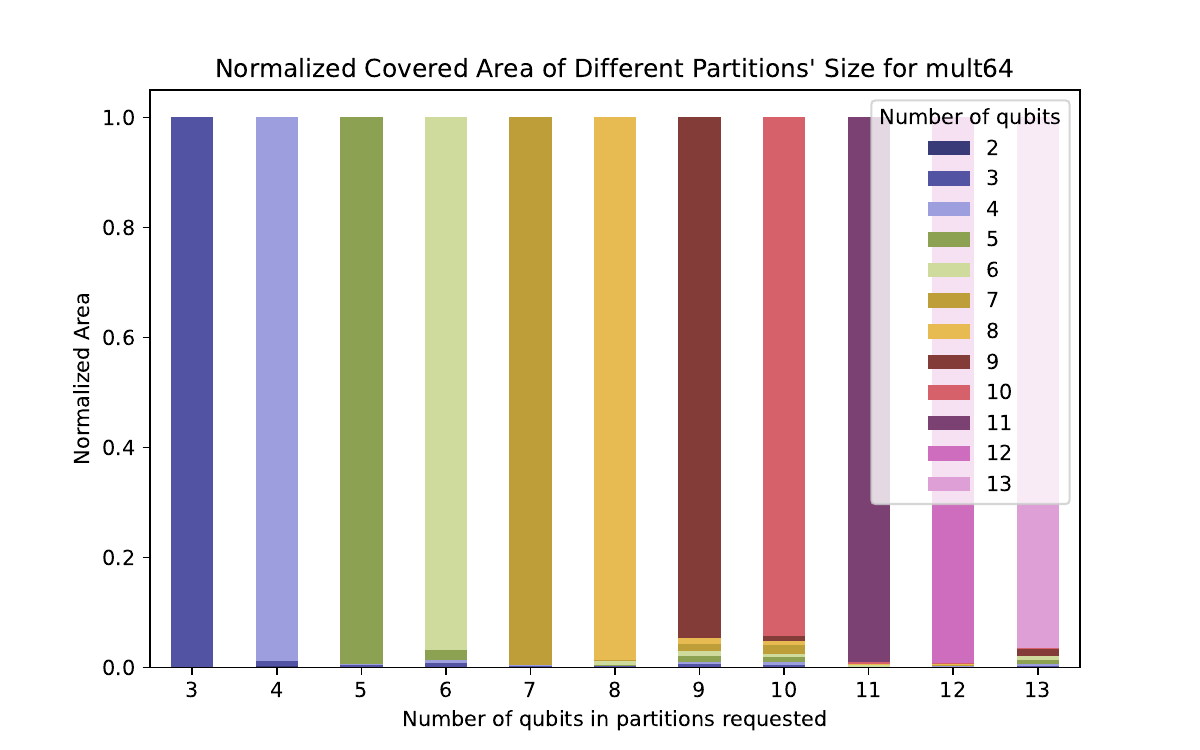}
    \includegraphics[width=0.445\linewidth]{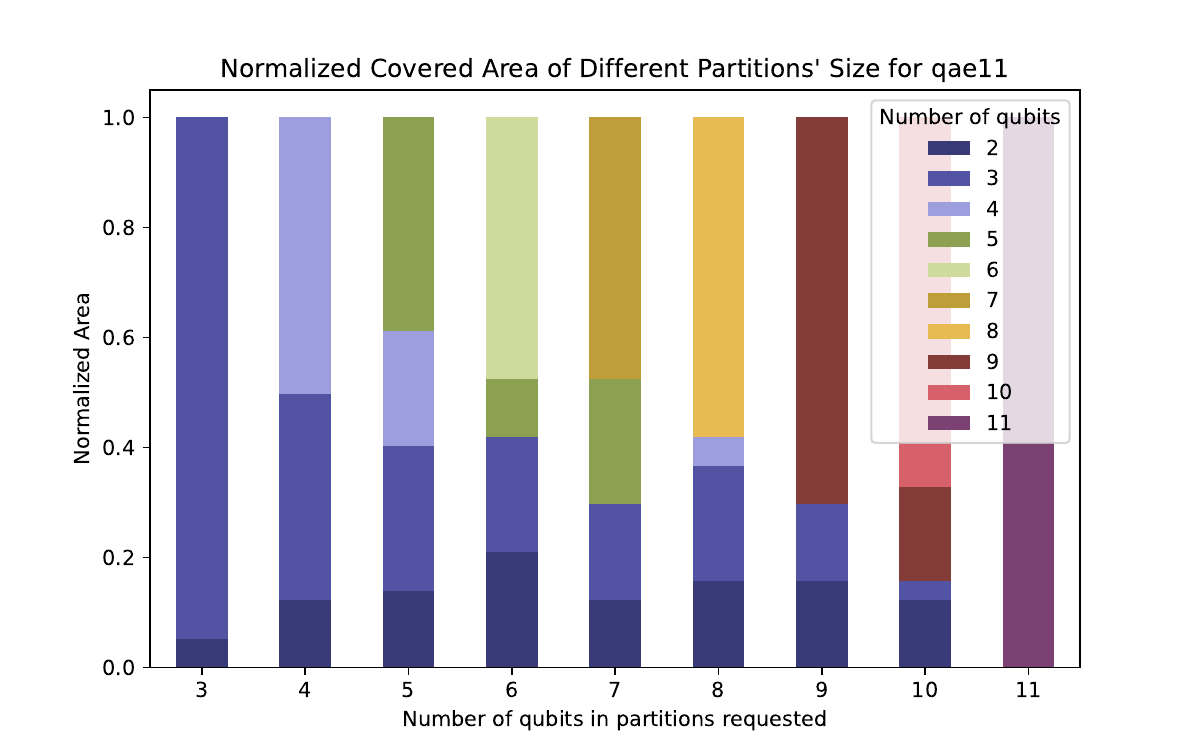}
    \caption{See caption on the next page.}
\end{figure}
\begin{figure}[htbp!]\ContinuedFloat
    \centering
    \includegraphics[width=0.445\linewidth]{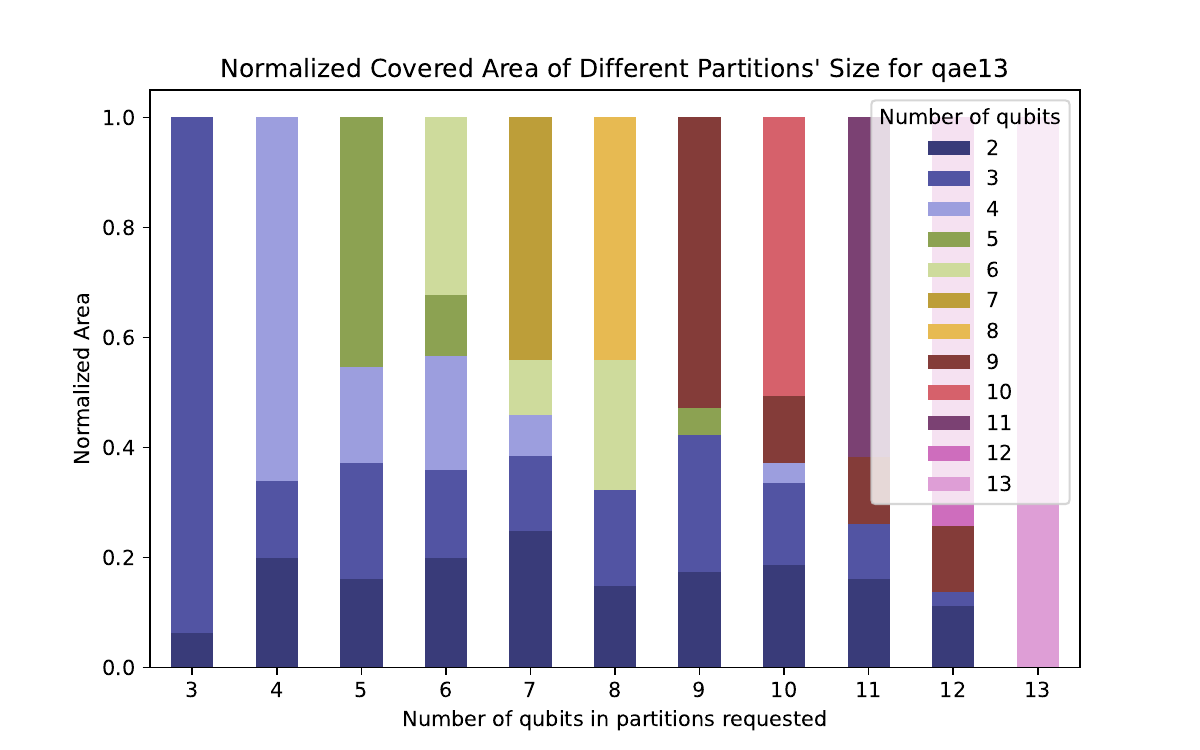}
    \includegraphics[width=0.445\linewidth]{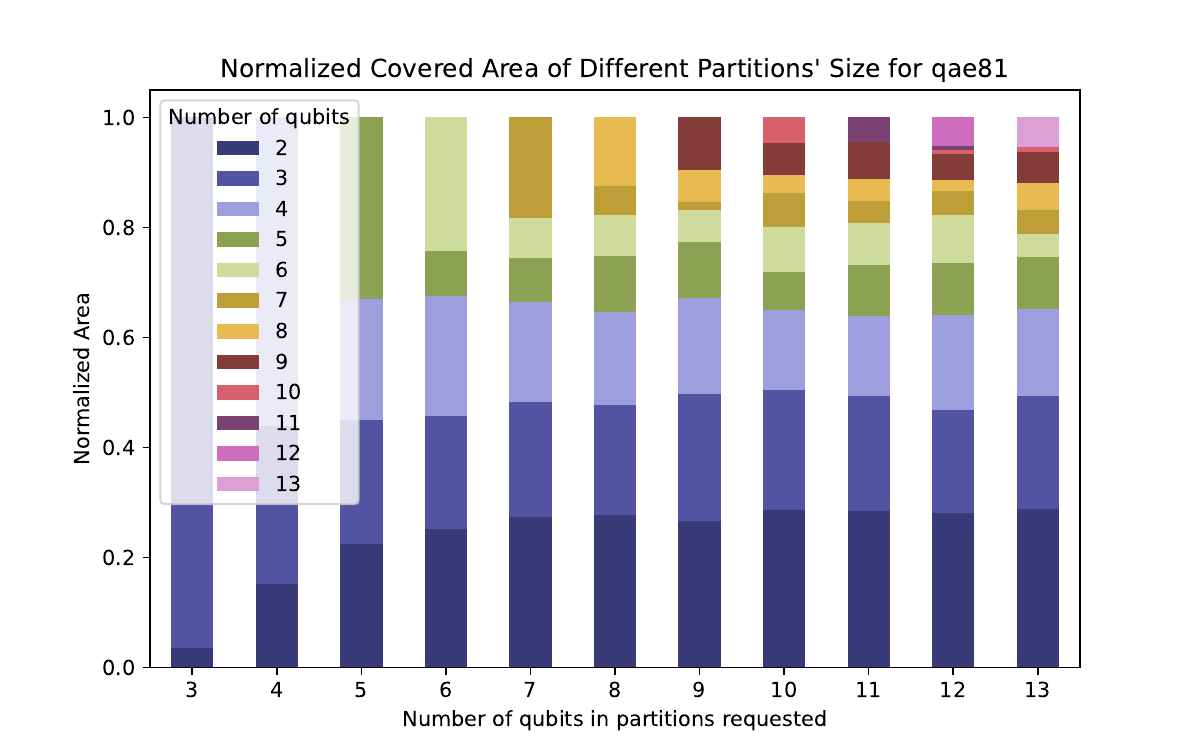}
    \includegraphics[width=0.445\linewidth]{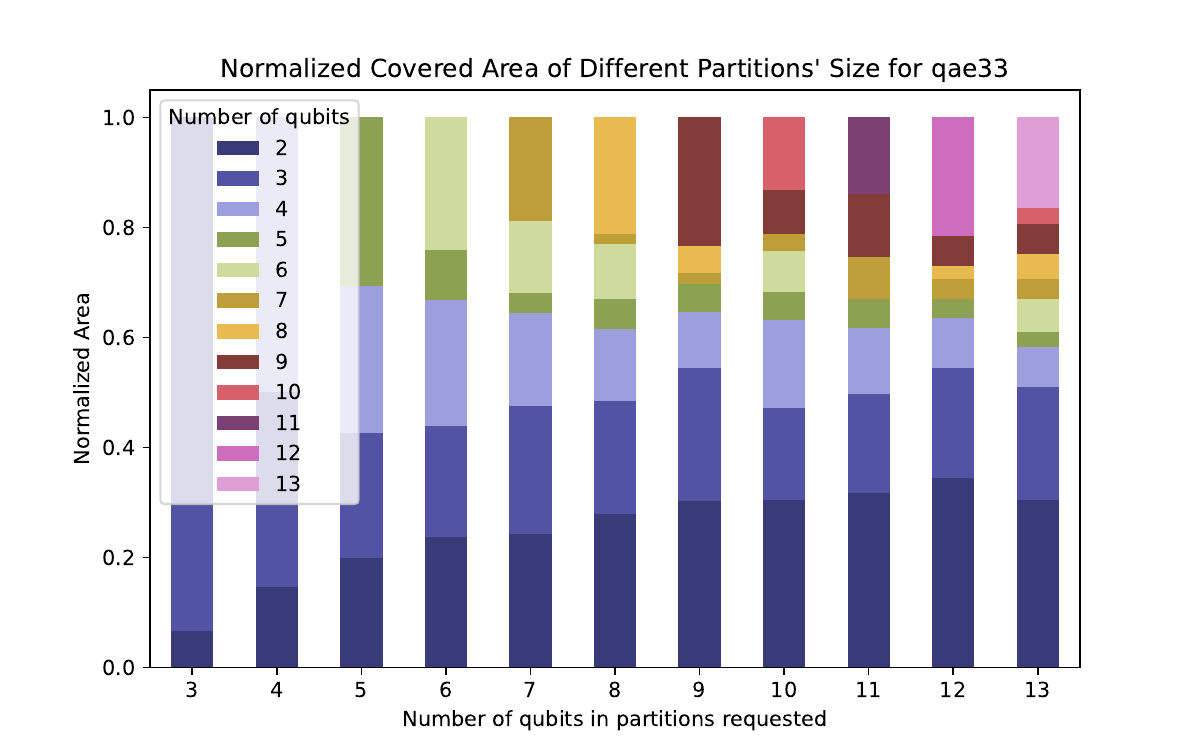}
    \includegraphics[width=0.445\linewidth]{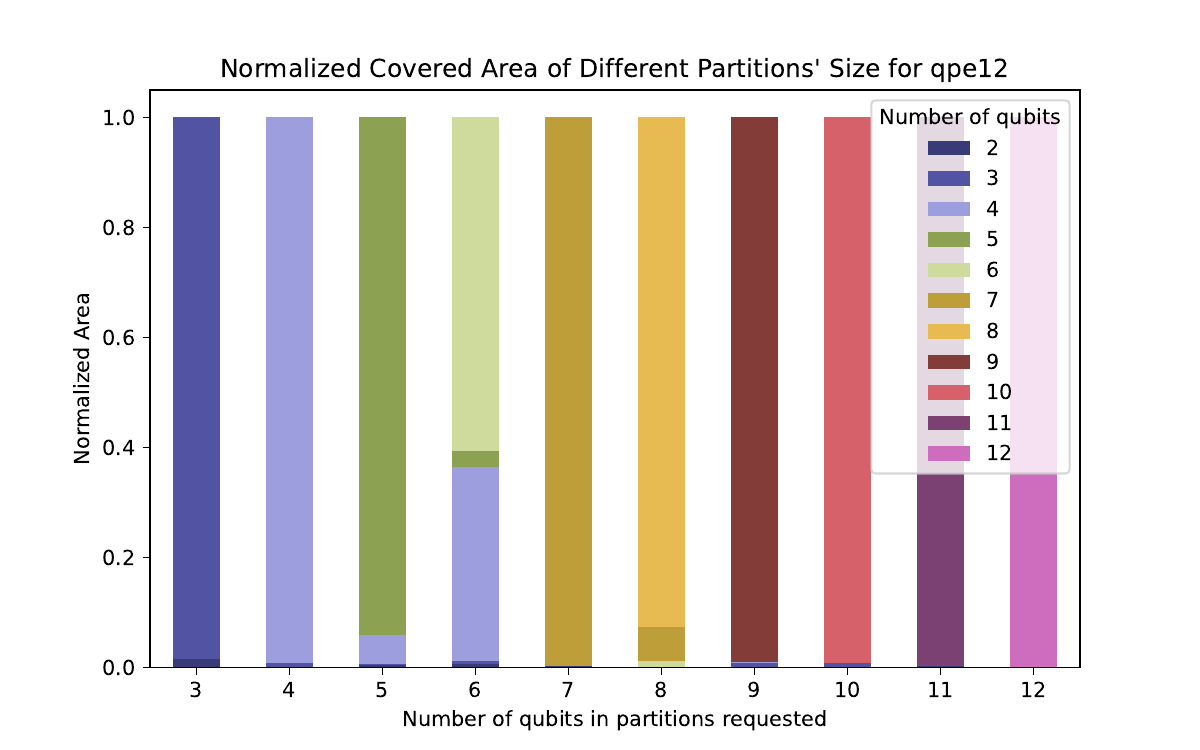}
    \includegraphics[width=0.445\linewidth]{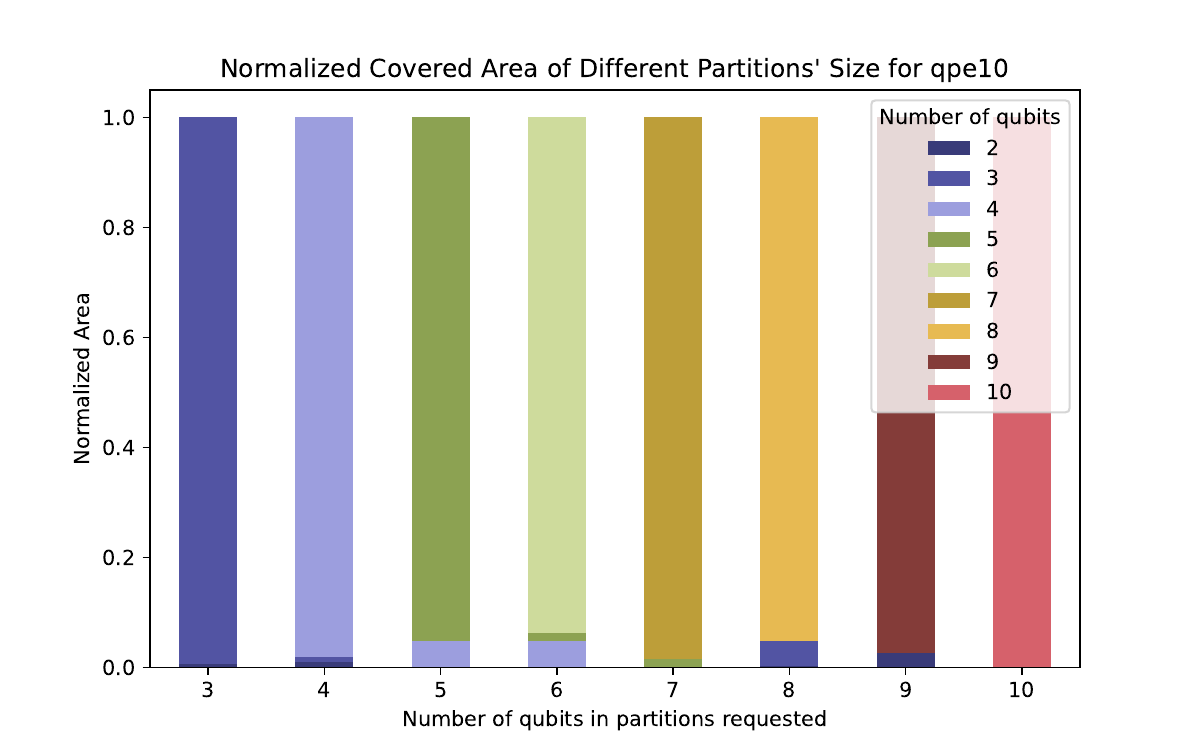}
    \includegraphics[width=0.445\linewidth]{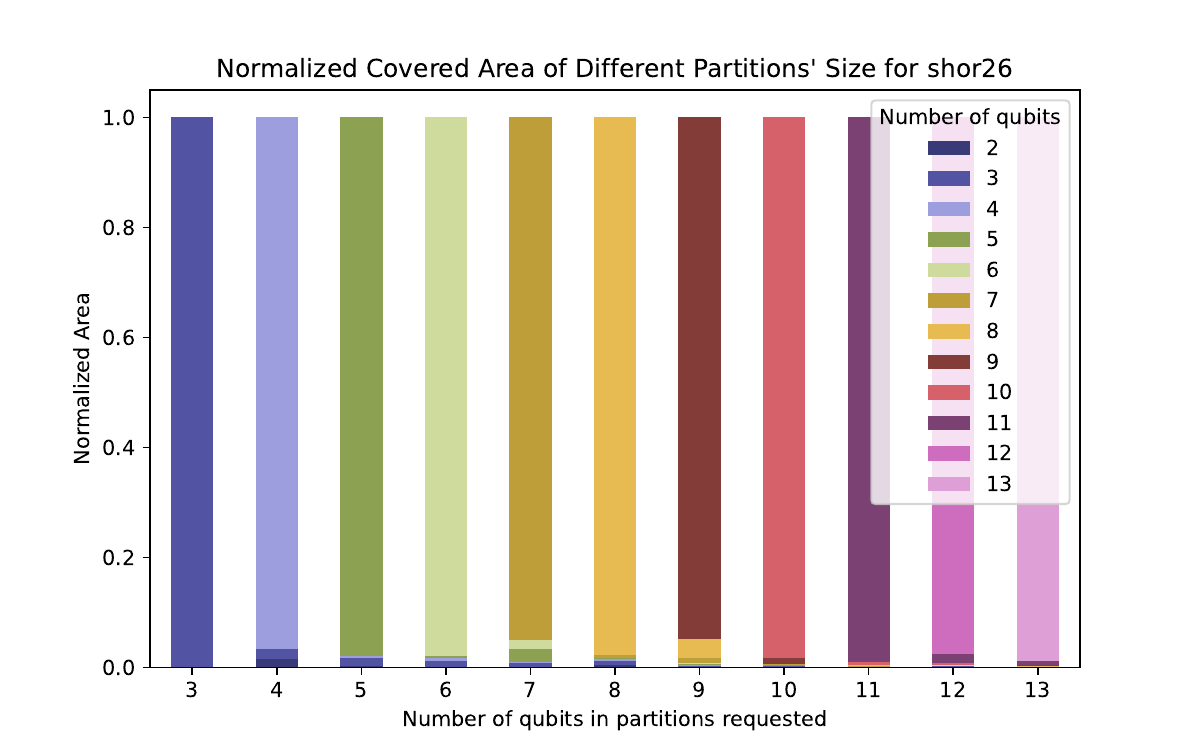}
    \includegraphics[width=0.445\linewidth]{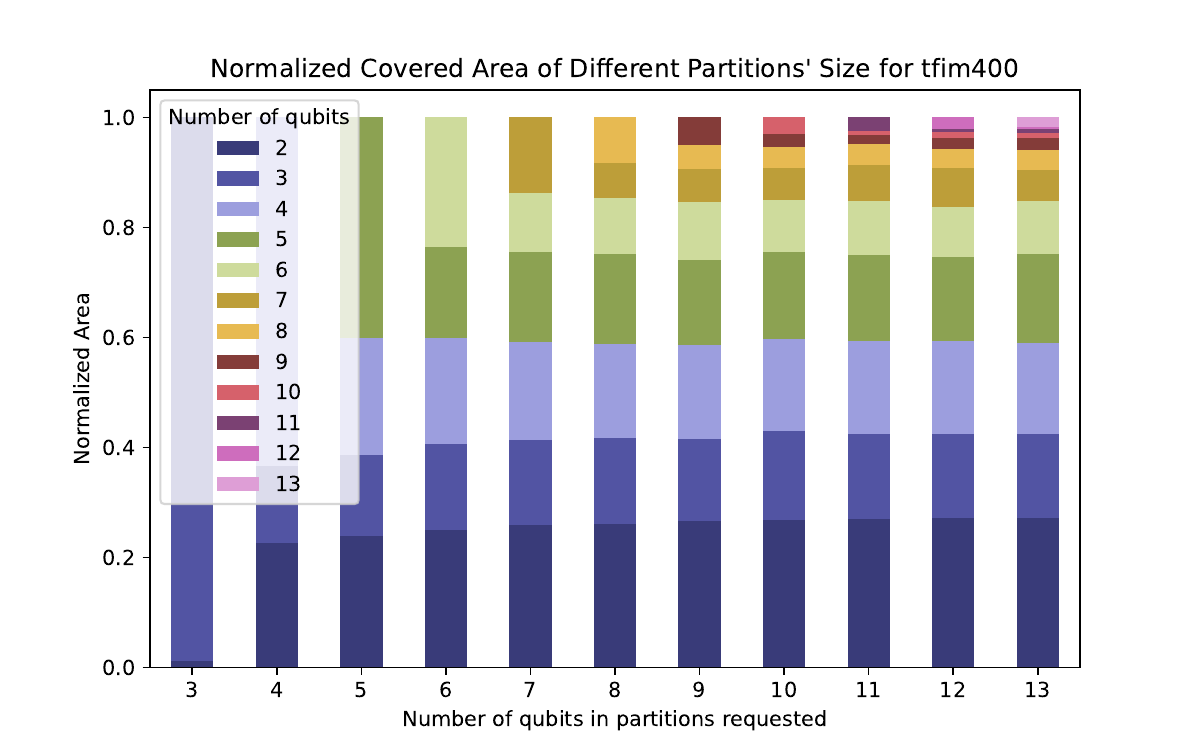}
    \includegraphics[width=0.445\linewidth]{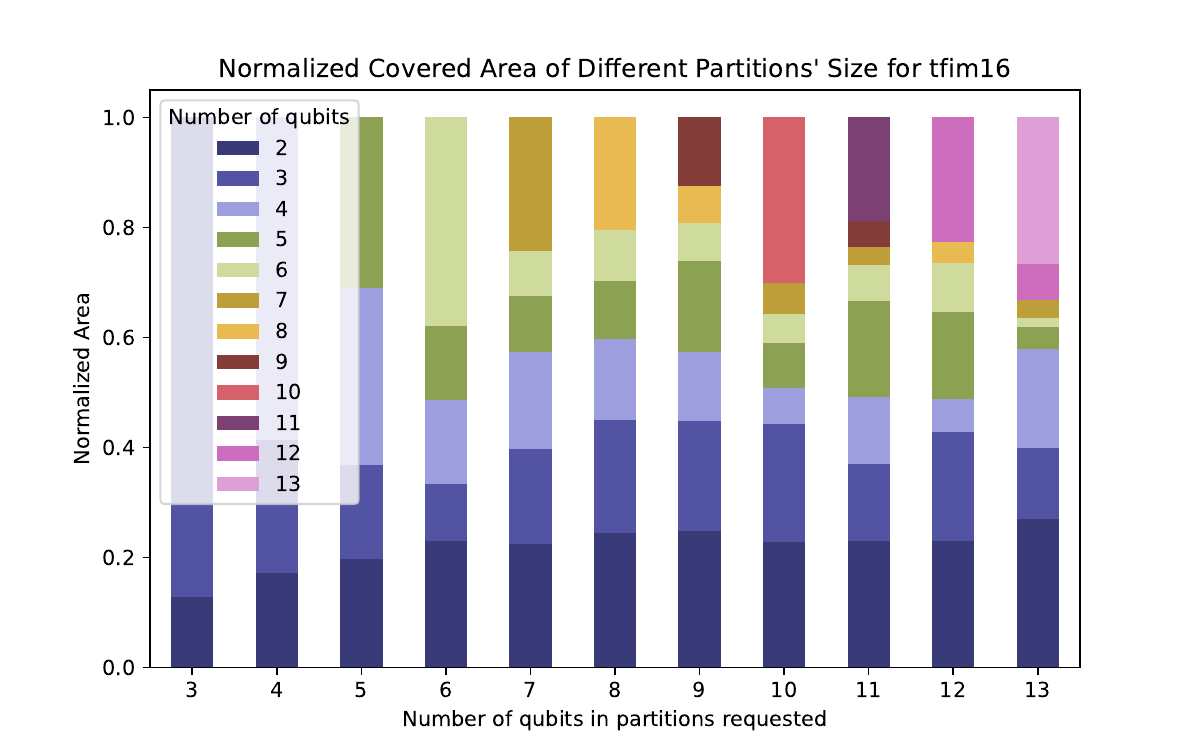}
    \includegraphics[width=0.445\linewidth]{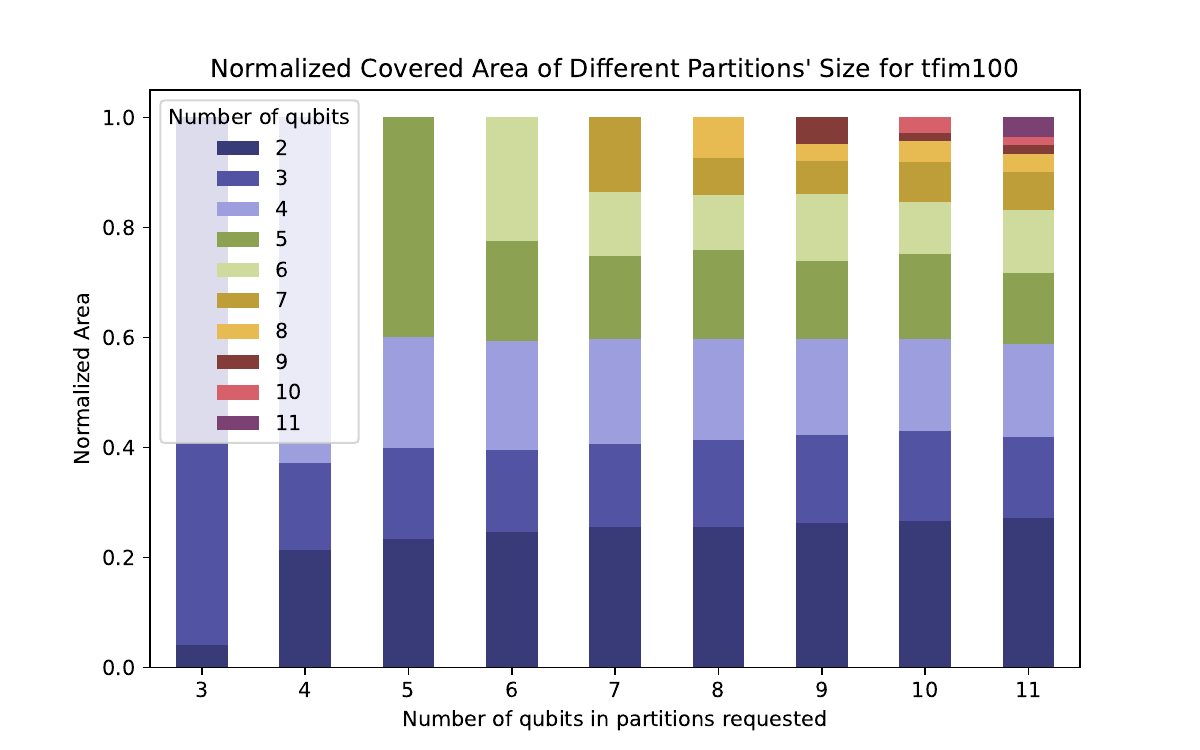}
    \includegraphics[width=0.445\linewidth]{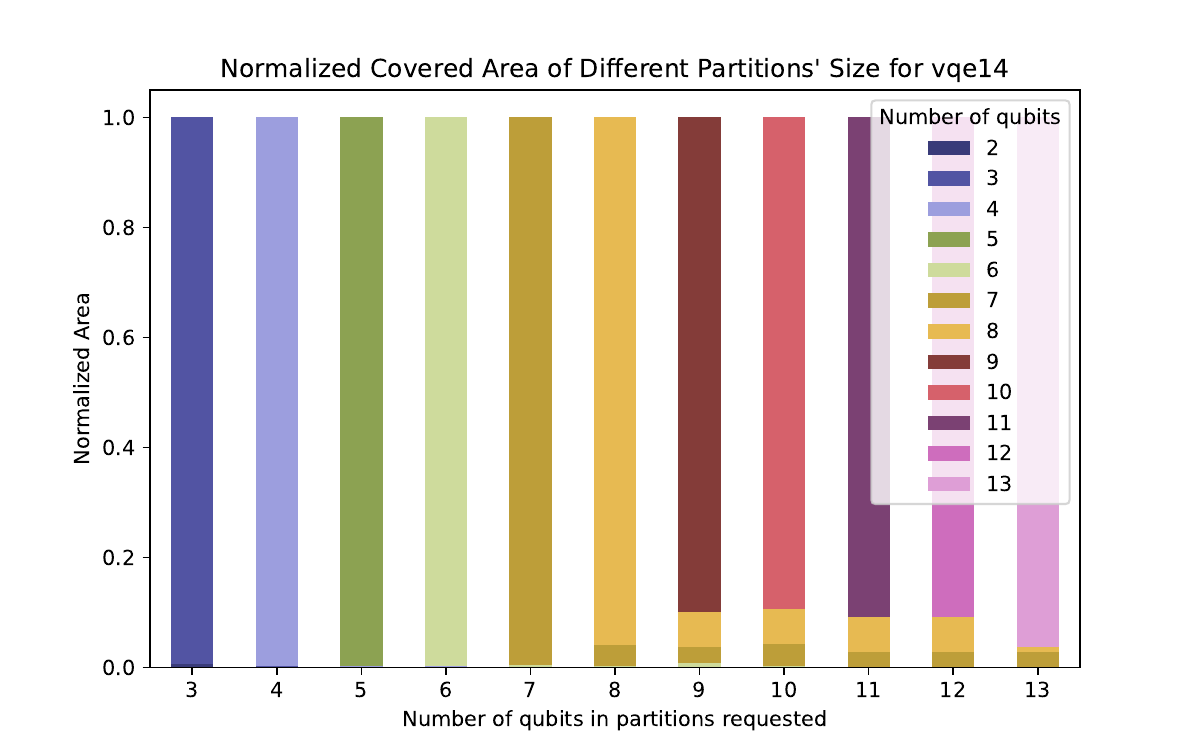}
    \caption{BQSkit's "QuickPartitioner" cover statistics for different circuits (See Table.~\ref{tab:benchmarks}) and partition sizes requested.}
    \label{fig:QuickPartitioner_stat}
\end{figure}
    

\section{Training States Distribution Impact}\label{app:tsd}

As shown in~\cite{caroGeneralizationQuantumMachine2022}, the generalization bounds are valid for arbitrary data-generating distribution. Some distributions will saturate that bound but some will perform better. As seen in practice: training based on computational basis states requires much more data than training on Haar random states. It is important to note that training in both cases satisfies the bound in~\eqref{eq:gen_bound}. All choices will lead to an exponential reduction in the time required to compute the cost function. This discussion is beyond the scope of the current paper though. It is because we divide the circuit into parts that can be classically simulated with a state vector simulator. Therefore, we can always afford to choose our data to be generated from Haar random distribution, which performed best in our tests. We have numerically verified that this choice leads to the smallest amount of data needed for good generalization.

We do not have a formal proof that the Haar distribution is optimal. However, we have numerical evidence that this is indeed the case as well as some intuitive argument that we describe below. Haar random distribution may not be universally optimal for every circuit but at least it avoids some problems other distributions face.

Let us say that our task is to compress a circuit $U$ that contains $R_z(\alpha)$ rotation as the first gate acting on the first qubit. Let us also assume that we are using computational basis states as our training data input states. The action of $U$ on any state of the form $\ket{0 b_1 b_2 \dots b_n}$ is insensitive to angle $\alpha$. That is, unless our training data set contains at least one state of the form $\ket{0 b_1 b_2 \dots b_n}$, it will be impossible to infer the full action of $U$ with this training data. That problem extends to other gates that act trivially on some computational basis states. Since we draw our training states randomly, this distribution choice will (on average) lead to an increased size of the training dataset needed to achieve generalization. Note that this problem is not present if we choose to work with Haar random input states. Up to a set of measure zero, each Haar random state is sensitive to angle $\alpha$ and will meaningfully contribute to probing the unitary $U$ with training states. One expects that the Haar random input state is likely to be sensitive to all parameters of the circuit $U$ and will lead to a small required size of the training dataset. Random input states do not introduce bias that may be present in other distributions as described above.

\end{acronym}
\end{document}